\documentclass[fleqn,usenatbib,twocolumn]{mnras}

\usepackage{newtxtext,newtxmath}
\usepackage[T1]{fontenc}
\usepackage{ae,aecompl}

\usepackage{etoolbox}
\makeatletter
\patchcmd\@combinedblfloats{\box\@outputbox}{\unvbox\@outputbox}{}{%
}%
\makeatother

\usepackage{graphicx}	% Including figure files
\usepackage{amsmath}	% Advanced maths commands
\usepackage{verbatim}
\usepackage{float}
\usepackage{academicons}
\usepackage{xcolor}
\usepackage{orcidlink}
\usepackage{hyperref}
\usepackage{color,soul}  % Highlight changes following referee's report
\usepackage{placeins}

\usepackage[a4paper]{geometry}
\newcommand{\orcid}[1]{\href{https://orcid.org/#1}{\textcolor[HTML]{A6CE39}{\aiOrcid}}}

\title
[Hydrodynamic simulations of flickering AGN jets]
{
Studying the link between radio galaxies and AGN fuelling with relativistic hydrodynamic simulations of flickering jets
}

\author[H. W. Whitehead, J.~H.~Matthews]{Henry W. Whitehead$^{1,2}$\thanks{henry.whitehead@physics.ox.ac.uk}\orcidlink{0009-0006-0716-0965} and James~H.~Matthews$^{1,2}$\thanks{james.matthews@physics.ox.ac.uk}\orcidlink{0000-0002-3493-7737} 
\\
$^{1}$Department of Physics, Astrophysics, University of Oxford, Denys Wilkinson Building, Keble Road, Oxford OX1 3RH, UK\\
$^2$Institute of Astronomy, University of Cambridge, Madingley Road, Cambridge, CB3 0HA, UK \\
}

\date{Accepted 2023 May 23. Received 2023 May 23; in original form 2023 April 19}

\pubyear{2023}

\begin{document}
\label{firstpage}
\pagerange{\pageref{firstpage}--\pageref{lastpage}}
\maketitle

\begin{abstract}
We present two- and three-dimensional hydrodynamic simulations of $\sim$kpc-scale AGN jets with mean jet powers in the range $1-7\times10^{45}\,$erg~s$^{-1}$, in which the jet power varies (through variation of the Lorentz factor) according to a flicker or pink noise power spectrum. We find the morphology and dynamics of the jet-cocoon system depends on the amplitude of the variability with a clear correspondence between the shape of the cocoon and the historical activity. The jet advances quickly during high-power states, whereas quiescent periods instead produce passive periods of inflation resembling Sedov-Taylor blast waves. Periods of high activity preferentially produce hotspots and create stronger backflow as they maximise the pressure gradient between the jet head and cocoon. The variability can also lead to propagating internal shock structures along the jet. Our work suggests that variability and flickering in the jet power has important implications, which we discuss, for observations of radio galaxies, ultrahigh energy cosmic ray acceleration and jet power to luminosity correlations. We explore the link between morphology and fuelling, and suggest that chaotic cold accretion should introduce a relatively small scatter in radio luminosity ($\sim0.2$\,dex) and modest imprints on morphology; sources such as Hercules A and Fornax A, which show evidence for more dramatic variability, may therefore require redder power spectra, or be triggered by mergers or other discrete events. We suggest ways to search for jet flickering observationally and propose that radio galaxies may be an important diagnostic of Myr timescale AGN fuelling, due to their `long-term memory'. 
\end{abstract}


%%%%%%%%%%%%%%%%%%%%%%%%%%%%%%%%%%%%%%%%%%%%%%%%%%

%%%%%%%%%%%%%%%%% BODY OF PAPER %%%%%%%%%%%%%%%%%%
\def\bb{\boldsymbol}
\def\alfven{Alfv\'en}
\def\alfvenic{Alfv\'enic}
\newcommand{\ergs}{erg~s$^{-1}$}
\newcommand{\tauesc}{\tau_{\rm esc}^{\rm cr}}
\newcommand{\pluto}{\textsc{Pluto}}
\def\gammajet{\Gamma_{\rm j}}
\def\adiabatic{\gamma_{\rm ad}}

\section{Introduction}
As Active Galactic Nuclei (AGN) accrete from their surroundings, they can expel material outwards in the form of jets and winds. Both these forms of outflow are important to our understanding of the accretion process, and each offer a medium through which the central black hole (BH) can influence proceedings far from its gravitational sphere of influence. Indeed, outflows are likely to be critical `AGN feedback' agents \citep[e.g.][]{fabian_observational_2012,morganti_many_2017,harrison_agn_2018,hardcastle_radio_2020}, potentially affecting star formation in the host galaxy, the fuelling of the AGN and the heating of the surrounding cluster or group environment. Collimated AGN jets can be generated from the magnetically driven extraction of rotational energy from either the accretion disc \citep{blandford_hydromagnetic_1982} or the black hole ergosphere \citep{blandford_electromagnetic_1977}. In both cases, the jet power is likely to be proportional to $\dot{M}c^2$ with some efficiency factor that depends on the detailed accretion physics and magnetic field topology. 

AGN jets produce a variety of observational signatures in various wavebands including X-rays and gamma-rays, but our focus here is primarily the radio emission observed on kpc to Mpc scales. Radio galaxies, which we take to mean AGN that emit strongly in the radio wavelengths \citep{hardcastle_radio_2020}, are characterised by strong synchrotron emission from nonthermal electrons gyrating in magnetic fields. These electrons are likely to be accelerated in locations where the jet dissipates kinetic energy, for example in shocks \citep[see][for a review]{matthews_particle_2020}. Radio galaxies are observationally characterised by a series of (blurry) dichotomies. Perhaps the most fundamental of these is the \citet[][FR]{fanaroff_morphology_1974} classification, which distinguishes radio galaxies based on their surface brightness distribution: an FR-I source is brighter towards the centre and dims at the edges, whereas an FR-II source is edge-brightened. These FR-II sources have bright hotspots at the end of their lobes, thought to be associated with the termination shocks of jets expelled by the AGN. The general paradigm is that FR-I sources are less powerful on average, but the picture is complicated by environment. Fundamentally, FR-IIs manage to remain collimated and thermalise much of their kinetic energy in a hotspot. On the other hand, FR-Is are not sufficiently powerful to avoid disruption and instead appear to gradually dissipate energy and decelerate, resulting in a plume-like emission structure. The relation between power and morphology is not straightforward; there are FR-II morphology sources at quite low luminosities \citep{mingo_revisiting_2019}, as well as a series of compact radio galaxies with diverse properties and varied nomenclature (e.g. FR0s [\citealt{baldi_fr0cat_2018}], compact symmetric and peaked spectrum sources [\citealt{odea_compact_1998,an_dynamic_2012}]).

Hydrodynamic (HD) simulations are a critical tool for studying how jets launch, propagate and dissipate energy \cite[see reviews by][]{davis_magnetohydrodynamic_2020,komissarov_numerical_2021}. For the simulation of jet launch, one must account for both general relativistic (GR) effects and magnetohydrodynamics (MHD). Over the past two decades, GRMHD simulations have been successful in producing jets from the Blandford-Znajek process \citep{mckinney_measurement_2004,tchekhovskoy_efficient_2011,mckinney_general_2012,porth_black_2017,liska_large-scale_2020}. On larger scales, it is common to use either HD and MHD approaches depending on the application. Early HD simulations were successful in producing the broad phenomenology of a jet beam that deposits energy in a shock, creates backflow, and inflates a cocoon while driving a bow shock into the surroundings \citep[e.g.][]{norman_structure_1982,falle_self-similar_1991}. There are a whole host of simulations of this type \citep[e.g.][]{duncan_simulations_1994,marti_morphology_1995,krause_very_2005,perucho_numerical_2007,mignone_high-resolution_2010,english_numerical_2016,perucho_long-term_2019}, now using up-to-date Godunov type finite volume codes, including detailed comparisons to observations, cluster `weather' and special relativistic effects. Many MHD simulations have also been conducted, particularly to explore the impact of magnetic kink instabilities \citep[e.g.][]{mizuno_three-dimensional_2009,mignone_high-resolution_2010,tchekhovskoy_three-dimensional_2016} and predict polarization signatures from the magnetic field structure \cite{hardcastle_numerical_2014}. A notable recent study from \cite{mukherjee_new_2020} includes ensembles of nonthermal electrons as tracer particles, accelerated at shocks in the simulation \citep{vaidya_particle_2018}, representing an advanced method for predicting the observational appearance of kpc-scale AGN jets \citep[see also][]{yates-jones_praise_2022,kundu2022,seo_simulation_2023}. 

A common approach to jet modelling is to inject something resembling a `top-hat' jet at the simulation boundary with a fixed jet power and thrust. This is clearly a reasonable approach to take without detailed knowledge of the multi-scale accretion process and history, but there is plenty of evidence that radio galaxy jets have variable powers. Double-double radio galaxies imply the existence of fairly discrete outbursts in powerful, classical radio sources \citep[e.g.][]{kaiser_radio_2000,konar_spectral_2006,konar_mode_2019}. Famous radio galaxies such as Centaurus A, Fornax A, Cygnus A and Hercules A also each show evidence of a dynamic history. For example, in Fornax A, \cite{maccagni_flickering_2020} find evidence for variable jet activity on $\sim 10$Myr timescales, with fresh activity near the nucleus appearing distinct from the episode that inflated the large radio lobes. Similarly complex behaviour can be seen in Centaurus A, whose giant ($300$kpc scale) lobes are disconnected from the current jet activity and $2$kpc scale inner lobes \citep{morganti_centaurus_1999,croston_high-energy_2009,hardcastle_high-energy_2009,wykes_mass_2013}. Hercules A shows bubble-like structures suggestive of sucessive explosive events \citep{dreher_rings_1984,gizani_multiband_2003,meier_heavy_1991,saxton_complex_2002,timmerman_origin_2021}. Cygnus A shows a classical double FRII morphology, but even this poster-child FRII source has multiple hotspots \citep{hargrave_observations_1974,carilli_cygnus_1996,stawarz_electron_2007,araudo_maximum_2018}. Clearly, morphologies of radio galaxies are complex, dynamic and far from steady or self-similar. This complexity can be shaped by a variety of processes including cluster/group environment, precession and projection effects, in addition to variation in the accretion rate and jet power; the latter is the focus of our study. 

Various authors have investigated the influence of variable jet powers, taking a variety of approaches. One method is to vary the power by turning the jet on and off in discrete bursts with a focus on dormant/dead or double-double radio galaxies \citep{reynolds_hydrodynamics_2002,mendygral_mhd_2012,yates_observability_2018,english_numerical_2019}. In an early and particularly relevant study, \cite{wilson_hotspots_1984} simulated a jet in which the velocity varied sinusoidally - this is possibly the first example of a hydrodynamic simulation of a variable jet. \cite{gomez_hydrodynamical_1997} also studied variability in jets by introducing perturbations in flow velocity with the aim of explaining superluminal radio sources. 
Several authors pursue a more self-consistent approach to study the feedback loop established between an AGN jet and the surrounding cluster or group environment, in which the accretion rate onto a region surrounding the BH is estimated and used to inform the power of the jet injected into the polar region \citep{yang_how_2016,beckmann_dense_2019,bourne_agn_2020,talbot_blandford-znajek_2021}. These studies generally support the idea of a feedback loop between the AGN jet and cluster or group environment, with a complex interplay between the jet power and both the backflow of material and heating/cooling of the surrounding medium. 

Recently, \cite{matthews_particle_2021} introduced a semi-analytic model for a radio galaxy with a flickering jet power and used it to study the particle populations accelerated by the jet. They adopted a flicker noise power spectrum (with the power spectral density [PSD] scaling as $1/f$ where $f $ is the temporal frequency), with a log-normal distribution of jet powers. This choice was motivated by simulations of AGN fuelling \citep{yang_how_2016,beckmann_dense_2019} as well as the `chaotic cold accretion' (CCA) model \citep{gaspari2013,gaspari2015,gaspari_self-regulated_2016,gaspari_raining_2017}. CCA is expected to produce a log-normal distribution of accretion rates with a flicker noise PSD. This set of statistical properties is also ubiquitous on shorter (human observable) timescales in accretion discs and is characteristic of a `multiplicative' physical process \citep[e.g.][]{lyubarskii_flicker_1997,uttley_flux-dependent_2001,uttley_non-linear_2005,gaskell_lognormal_2004,scaringi_physical_2014,alston_remarkable_2019,alston_non-stationary_2019}. In a more generic sense, flicker noise is also extremely common in a whole host of settings: in music, electronic circuits, and other astrophysical environments \citep{press_flicker_1978}. Thus, while the characteristics of the Myr-timescale variability in radio galaxies are unknown and debatable, it seems reasonable to think of jet power variation as a {\em noisy} process. We can then investigate the impact this noise might have on the structures and shapes we observe as well as the dynamics of the jet-cocoon system. 

In this work -- and with the above astrophysical context in mind -- we conduct relativistic hydrodynamic simulations of kiloparsec-scale jets in which the input jet power is varied systematically according to a flicker noise, or pink noise spectrum. This approach allows to study the morphology of the jet produced and how it responds to the variation of input power. Our simulations are broadly designed to mimic the conditions in moderately powerful radio galaxies on the cusp of the traditional (and somewhat outdated) FRI-FRII power divide. In addition, our relativistic treatment allows us to investigate the impact of varying Lorentz factor at the jet inlet. Our paper is structured as follows. We begin by describing the numerical method and simulation grid we use in section~\ref{sec:method}. In section~\ref{sec:2d_results}, we describe results from a small grid of 2D, axisymmetric simulations using various random number seeds and variability parameters, before presenting a fiducial 3D simulation in section~\ref{sec:3d_results}. We discuss our results in section~\ref{sec:discuss}, including the limitations of our work, before concluding in section~\ref{sec:conclusions}. 

\section{Numerical Method}
\label{sec:method}

\subsection{Simulation Setup}
We use the publicly available code \pluto\ (version 4.3) \citep{mignone_pluto_2007} to solve the equations of Relativistic Hydrodynamics (RHD). We neglect magnetic fields (see section~\ref{sec:limitations}). We use linear reconstruction with characteristic limiting, second order Runge-Kutta time stepping and a dimensionally unsplit scheme. We use the Harten-Lax-van Leer-Contact (HLLC) Riemann solver. Broadly speaking, our setup follows that of \cite{matthews_ultrahigh_2019}. Our models require a relativistic treatment as during high power phases the bulk Lorentz factor $\gammajet$ of the jet can reach $\approx15$. The 2D simulations are conducted in cylindrical geometry with axisymmetry about $x=0$, whereas the 3D simulations use a regular cartesian grid.  

\pluto\ solves continuity equations in the form
\begin{equation}
    \frac{\partial \textbf{U}}{\partial t} = - \nabla \cdot \textbf{T(U)} + \textbf{S(U)},
\end{equation}
where \textbf{U} is a vector of conserved quantities, the fluxes for the components of which form the rows of \textbf{T}, and \textbf{S} denotes source terms. For the RHD module, \textbf{U} and \textbf{T} are given by
\begin{equation}
    \textbf{U} = \begin{pmatrix} D \\ \bf{m} \\E \end{pmatrix} \quad \quad \textbf{T} = \begin{pmatrix} D\bf{v} \\ \textbf{mv} + P\textbf{I} \\ \textbf{m} \end{pmatrix}^T
\end{equation}
where $D$ is the laboratory density, \textbf{m} is the 3-momentum, E is the total energy, \textbf{v} is the 3-velocity and $p$ is the pressure. The conserved quantities in \textbf{U} are related to the primitive variables $\rho$, $P$ and $\textbf{v}$ through
\begin{equation}
    D = \rho\Gamma, \quad \textbf{m} = \rho h \Gamma^2 \textbf{v}, \quad E = \rho h \Gamma^2 - P,
    \label{eq:conserved}
\end{equation}
there we have introduced the specific enthalpy $h \equiv c^2 + P/\rho + e$, where $e$ is the specific internal energy. We use the Taub-Mathews equation of state \citep{taub_relativistic_1948,mignone_equation_2007} to describe a relativistic plasma. This equation of state allows us to smoothly transition between relativistic and non-relativistic temperatures. The adiabatic index for our material will be consistent with a nonrelativistic ideal gas (adiabatic index $\adiabatic$ = 5/3) at low temperatures and a relativistic plasma ($\adiabatic$ = 4/3) at high temperatures. 

Our simulation domain for each 2D model is an axisymmetric mesh of $640\times1536$ cells corresponding physically to $125\times300$\,kpc, giving a resolution of $0.195$\,kpc. In 3D we use a slightly shorter domain in the $z$ direction of $157.5\times157.5\times 200$\,kpc with $504\times504\times640$ cells, at a slightly lower resolution of $0.3125$\,kpc. In \pluto, physical values are converted to simulation units to avoid handling extremely large or small numbers during calculation. We adopt a simulation unit length ${\cal L}_0 = 3.086 \times 10^{21}$\,cm ($1$\,kpc) and density $\rho_0 = 6 \times 10^{-27}$ g cm$^{-3}$. With the requirement for RHD that the unit velocity be $c$, we can define the other derived unit dimensions of pressure $P_0 = \rho_0 c^2 = 5.393 \times 10^{-6}$ dyne cm$^{-2}$ and time $t_0 = {\cal L}_0/c = 3.264 \times 10^{3}$\,yr. 

\subsubsection{Jet injection}
Our jet material is injected at the origin with velocity $\bf{v_{\rm j}}$ = $v_{\rm j} \bf{\hat{e}_z}$ and rest frame density $\rho_{\rm j} = \rho_0 \eta$, where $\eta$ is the density contrast. For the models discussed here, $\eta = 10^{-4}$, meaning the jet material is significantly less dense than the ambient medium in its rest frame. The material is highly supersonic, with the initial internal Mach Number ${\cal M}_{{\rm j},0} = 100$. However, the jet undergoes a series of reconfinement shocks along its length and so ${\cal M}_{\rm j}(z)$ further along the jet is rather insensitive to this initial value. With a jet inlet radius of 1 kpc, our jet radius is reasonably well resolved by 5 cells in the 2D simulations and just over 3 cells in the 3D simulations. \cite{english_numerical_2019} found that 2.7 cells covering the jet radius was sufficient, so we expect our resolution across this region to effectively capture the injection of energy and momentum by the jet. While the jet quantities are nominally uniform across the inlet, they are smoothed slightly to avoid discontinuities at the inlet edge, such that any jet quantity (velocity, density, pressure) $\xi_{\rm j}$ varies on the $z = 0$ boundary as $\xi_{\rm j}(x,z=0) = \xi_{\text{inlet}}/\cosh(x^{14})$. As such the $z = 0$ boundary condition is inflow for $x<r_{\rm j}$ (jet inlet) and reflective outside of this. The outer $x,y$ and $z$ boundaries have outflow boundary conditions and, in 2D, the $x = 0$ boundary condition is axisymmetric.

\begin{figure}
    \centering
    \includegraphics[width=0.8\linewidth]{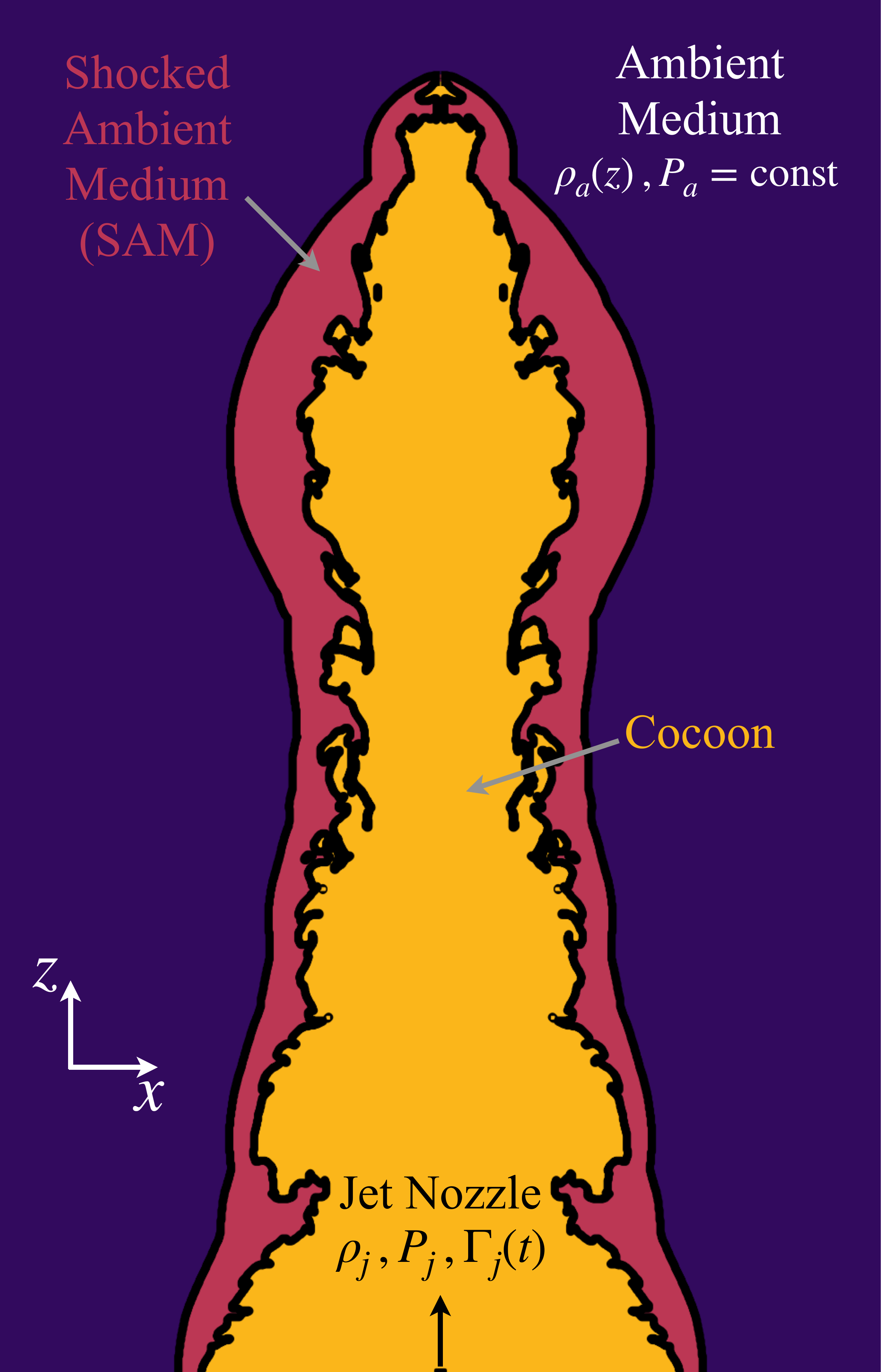}
    \caption{Diagram showing the basic set-up of the simulation and the identification of the different zones described in the text (cocoon, SAM, ambient medium, shown in different colours) for one of the snapshots shown later in Fig.~\ref{fig:4panel_sigma1.5}.}
    \label{fig:regions}
\end{figure}

\subsubsection{Cocoon and bow shock identification}
\label{sec:regions}
We distinguish the jet cocoon, shocked ambient medium and surrounding atmosphere using a combination of the pressure and the presence or absence of jet material. To track jet material, we include a standard jet tracer, $C_{\rm j}$, which, at the $z=0$ boundary, is set to 1 inside the jet inlet and 0 everywhere else. This tracer is evolved as a passive scalar according to the equation
\begin{equation}
    \frac{\partial (\rho C_{\rm j})}{\partial t} = - \nabla \cdot (\rho C_{\rm j} \boldsymbol{v}).
\end{equation}
Mixing of jet material with the background will dilute this quantity. Domain cells that have a tracer value of greater than $10^{-4}$ are considered majority jet material and so are labelled as within the jet cocoon (see also section~\ref{sec:2d_dynamics}). The bow shock is identified by searching inwards for a doubling in pressure. Material enclosed by the bow shock but outside of the jet cocoon is labelled as shocked ambient material. We checked visually that both our approaches were effective at identifying the bow shock and contact discontinuity, and an example is demonstrated in Fig.~\ref{fig:regions}. The background medium remains undisturbed until the bow shock passes through it, and is not relevant for this investigation. 

\subsubsection{Ambient medium}
To simulate the medium surrounding an AGN, we set the background density using an isothermal `beta' model or King profile with core radius $r_c$ = 50 kpc and $\beta$ = 0.5 such that
\begin{equation}
    \rho (r) = \rho_0 \left[ 1 + \left(\frac{r}{r_c}\right)^2\right]^{-\frac{3\beta}{2}}.
\end{equation}
These adopted values of $r_c$ and $\beta$ are fairly typical for the environments of radio-loud AGN studied by \cite{ineson_link_2015}. 
In contrast to some other studies, we use a uniform pressure background with $P_{\rm a} = 6\times10^{-9}~P_0  = 3.24\times10^{-14}{\rm dyne~cm}^{-2}$. A uniform pressure is not physically realistic, because the pressure decreases with radius in group and cluster environments \citep[e.g.][]{arnaud_universal_2010,sun_pressure_2011}. However, in the \pluto\ RHD module the gravitational potential in the momentum equation is not well-defined, and any pressure gradient in the atmosphere must be balanced by a gravitational potential to keep the atmosphere stable. In practice, our choice here does not make much difference to the results -- we found that runs with and without a realistic pressure gradient and balancing gravitational force produced very similar structures and morphologies, albeit with differences in advance speed at the $\approx10$\% level. 

\subsection{Synthetic Jet Power Time Series}
To model jet variability, we follow \cite{matthews_particle_2021} in assuming that the jet power varies according to a time series with certain pre-specified statistical properties. To generate the time series, we use the method described by \cite{emmanoulopoulos_generating_2013} and implemented in Python by \cite{connolly_python_2015} to generate a series of powers $Q(t)$. To properly characterise our spectrum we require $p(Q)$, the probability density function (PDF) and $S(f)$, the power spectral density. The parameters of this time series -- which control the variability of the simulation -- are:
\begin{itemize}
    \item Median Jet Power: $Q_0 = 10^{45}~{\rm erg s}^{-1}$
    \item PSD frequency index: $\alpha_{f}$ = 1
    \item Variability: $\sigma$ (free parameter)
    \item Random number seed: $i$ (free parameter)
\end{itemize}
The median power, $Q_0$, could take many values as, observationally, jet power spans a wide range. We select a median power close the canonical FRI-FRII divide and comparable to the average jet power reported by \cite{yang_how_2016}. We keep $Q_0$ fixed both to restrict our parameter space and because we are interested mostly in the relative impact of the variability parameter $\sigma$.

The PSD defines the magnitude of frequency components in our light curve, as follows:
\begin{equation}
    S(f) \propto f^{-\alpha_{f}}
\end{equation}
In setting $\alpha_f = 1$ we adopt a pink/flicker noise power source, motivated by the chaotic cold accretion model \citep{gaspari_self-regulated_2016} and the general fuelling arguments presented in the introduction. Based on the same studies, we use a log-normal distribution of jet powers, so the jet power PDF $p(Q)$ obeys 
\begin{equation}
    p(Q) = \frac{1}{\sqrt{2\pi}\sigma Q}\exp\left[-\frac{(\ln Q-\ln Q_0)^2}{2\sigma^2}\right].
\end{equation}
Our use of a log-normal distribution means that the moments of the PDF differ from that of a standard Gaussian. \cite{matthews_particle_2021} discuss the behaviour of log-normal distributions further in the context of jets and particle acceleration, specifically exploring how the mean $\overline{Q}$ and mode Mo$(Q)$ diverge from the median with increasing $\sigma$, according to the equations
% Here $\mu$ is the logarithmic median ln($Q_0$).
\begin{equation}
    \overline{Q}(\sigma) = Q_0 \exp\left(\frac{\sigma^2}{2}\right) \quad \quad \text{Mo}(Q) = Q_0 \exp\left(-\sigma^2\right).
\label{eq:mean}
\end{equation}
In our grid of models we vary $\sigma$ between 0 and 1.5, so as to compare the effects of variability on the jet's evolution. The other controlling parameter we use to distinguish simulations is the random seed. 

From our jet power series, we can assign a jet power at each time step of the simulation. In relativistic hydrodynamics  \citep[e.g.][]{taub_relativistic_1948,landau_classical_1975,wykes_1d_2019}, the jet power is given by 
\begin{equation}
    Q = \pi r^2_{\rm j} v_{\rm j} \left[\gammajet\left(\gammajet - 1\right)\rho_{\rm j} c^2 + \frac{\gamma_{\rm ad}}{\adiabatic - 1}\gammajet^2 P_{\rm j}\right]
\end{equation}
where $\adiabatic$ is the adiabatic index, $\gammajet$ is the jet inlet bulk Lorentz factor and $v_{\rm j}$, $\rho_{\rm j}$, $P_{\rm j}$ and $r_{\rm j}$ are the jet beam velocity, density, pressure and jet width (1 kpc). This equation technically holds only if the jet quantities are homogeneous across the inlet, but our smoothing function is suitably steep for this to be a reasonable approximation. As our jet beam is highly supersonic, we  neglect the energetic contributions from the pressure, and solve for $\gammajet$ numerically by inverting the above equation. Thus we are able to pre-generate time series containing the proper $\gammajet$ values provided the simulation parameters $\eta$, $\rho_0$ and $r_{\rm j}$ are specified. This time series is used by \pluto\ at each time step to set the jet inlet boundary condition. 

\begin{figure}
    \centering
    \includegraphics[width=\linewidth]{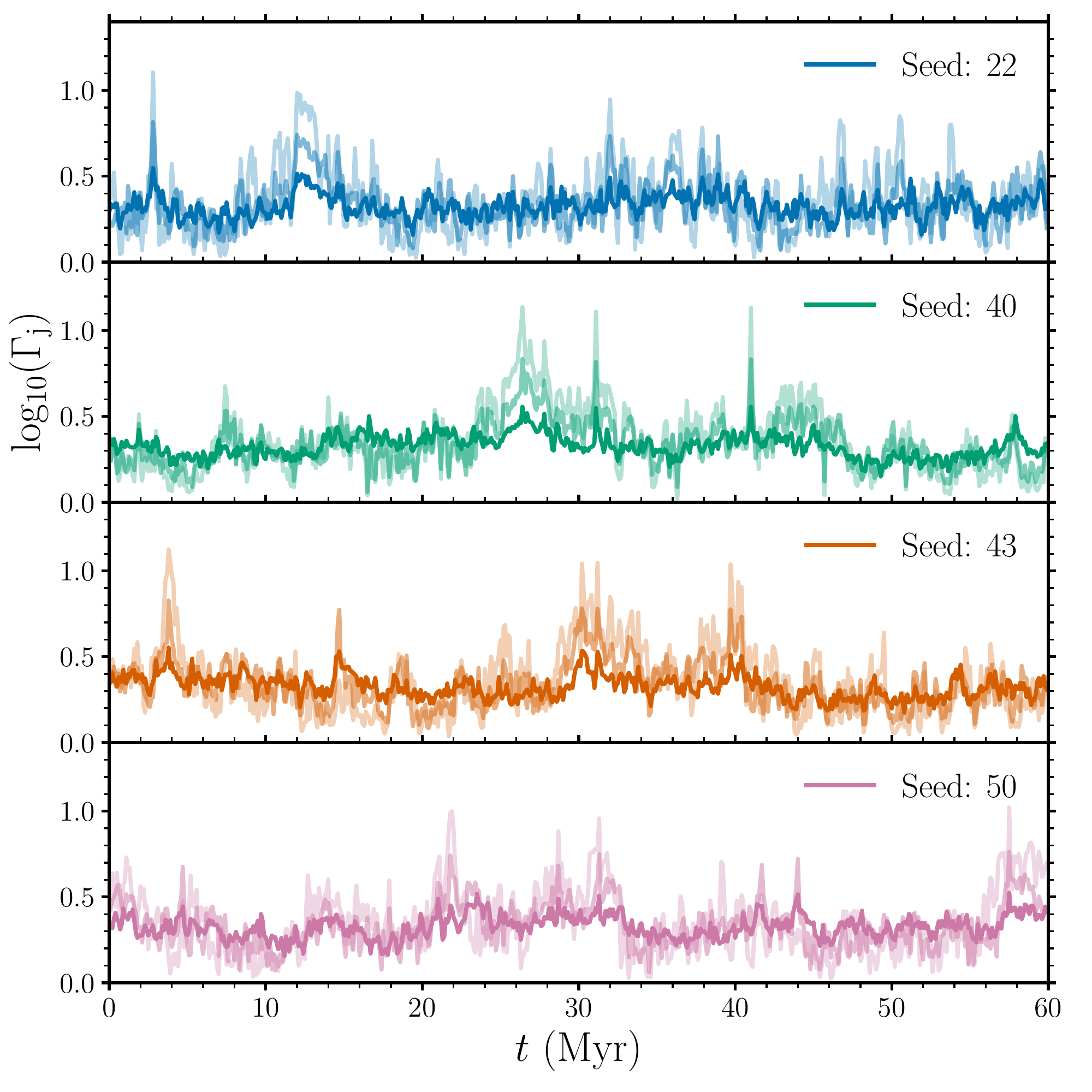}
    \caption{Synthetic jet bulk Lorentz factor time series used in this work, for the four different random number seeds and values of $\sigma \in (0.5,1,1.5)$. Each panel shows a given seed, with the opacity of the line decreasing for increasing $\sigma$.}
    \label{fig:power_series}
\end{figure}

\begin{table}
\centering 
\renewcommand{\arraystretch}{1}
\begin{tabular}{c c c}
\hline
    Parameter & Description & Value(s) \\ \hline \hline 
    \multicolumn{3}{l}{Grid Points} \\ 
    \hline 
    $i$ & Random number seed &  $\{22,40,43,50\}$ \\
    $\sigma$ & Variability parameter &  $\{0,0.5,1,1.5\}$ \\
    \hline 
    \multicolumn{3}{l}{Fixed Parameters} \\ 
    \hline 
    $\eta$ & Jet Density Contrast ($\rho_{\rm j}/\rho_0)$  & $10^{-4}$ \\ 
    $r_{\rm j}$ & Jet Inlet Radius & $1$\,kpc \\
    $\beta$ & Density profile exponent & $0.5$ \\
    $r_c$ & Density profile core radius & $50$\,kpc \\
    $P_{\rm a}$ & Ambient pressure (${\rm dyne~cm}^{-2}$) &  $3.24\times10^{-14}$ \\
    ${\cal M}_{\rm j}$ & Internal Mach number & $100$ \\
    $\alpha_f$ & Variability PSD slope & $1$ \\ 
    \hline 
    \multicolumn{3}{l}{Derived Quantities} \\ 
    \hline
    $\overline{Q}_{\rm j}$ & Mean jet power &  $10^{45}\times e^{\sigma^2/2}$\,erg~s$^{-1}$ \\
    $Q_0$ & Median jet power & $10^{45}$\,erg~s$^{-1}$ \\
    \hline
\hline
\end{tabular}
\caption{
Parameters for the simulation grid.
}
\label{table:grid}
\end{table}

\subsection{Simulation Grid}
Our simulation grid comprises 16 models in 2D and one fiducial 3D model, with the results described in sections~\ref{sec:2d_results} and \ref{sec:3d_results}, respectively. Each variable jet model is described by two free parameters, $\sigma$ and the random number seed $i$. To avoid time series with very large spikes which could occur very early or late in the simulation, we restricted our models by requiring that, for the highest $\sigma$, the bulk Lorentz factor did not exceed $15$, and that the cumulative energy $E=\int Q(t) dt$ deposited in the first or last 30 Myr of the time series did not exceed 60\% of the total energy. Selecting the first 4 seeds to meet this criteria, we used random number seeds 22, 40, 43 and 50 for our investigation. With these choices, our simulation grid is defined as follows: 
\begin{itemize}
    \item 12 variable models spanning 
    $$
    \sigma \in \{0.5,1.0,1.5\},~
    i \in \{22,40,43,50\} \nonumber
    $$
    \item 1 model with constant power $Q_0$ (the median power)
    \item 3 models with constant power $\overline{Q}(\sigma)$, the mean power for each value of $\sigma$ used in the variable models 
\end{itemize}
The random number seed $i$ only applies to variable models. The time evolution of the input Lorentz factor in each of the 12 variable models is displayed in Figure \ref{fig:power_series}, with decreasing opacity indicating increasing variability. All of the simulations were run on the University of Cambridge's high performance computing system, using the Peta4 supercomputer run by the Cambridge Service for Data-Driven Discovery (CSD3). Each 2D model was parallelised (using {\sl Message Parsing Interface} [MPI] parallelisation) across 64 cores and ran for $\approx90-120$ minutes, whereas the 3D model was parallelised across 504 cores and run for $\approx12$~hours ($\approx6048$ core hours). 

\section{Axisymmetric 2D Results} 
\label{sec:2d_results}

We begin by describing results from our small grid of axisymmetric 2D simulations, focusing particularly on how the dynamics and morphology of the jet-cocoon system varies for different values of $\sigma$ and different random number seeds. 

\begin{figure*}
    \centering
    \includegraphics[width=1.0\linewidth]{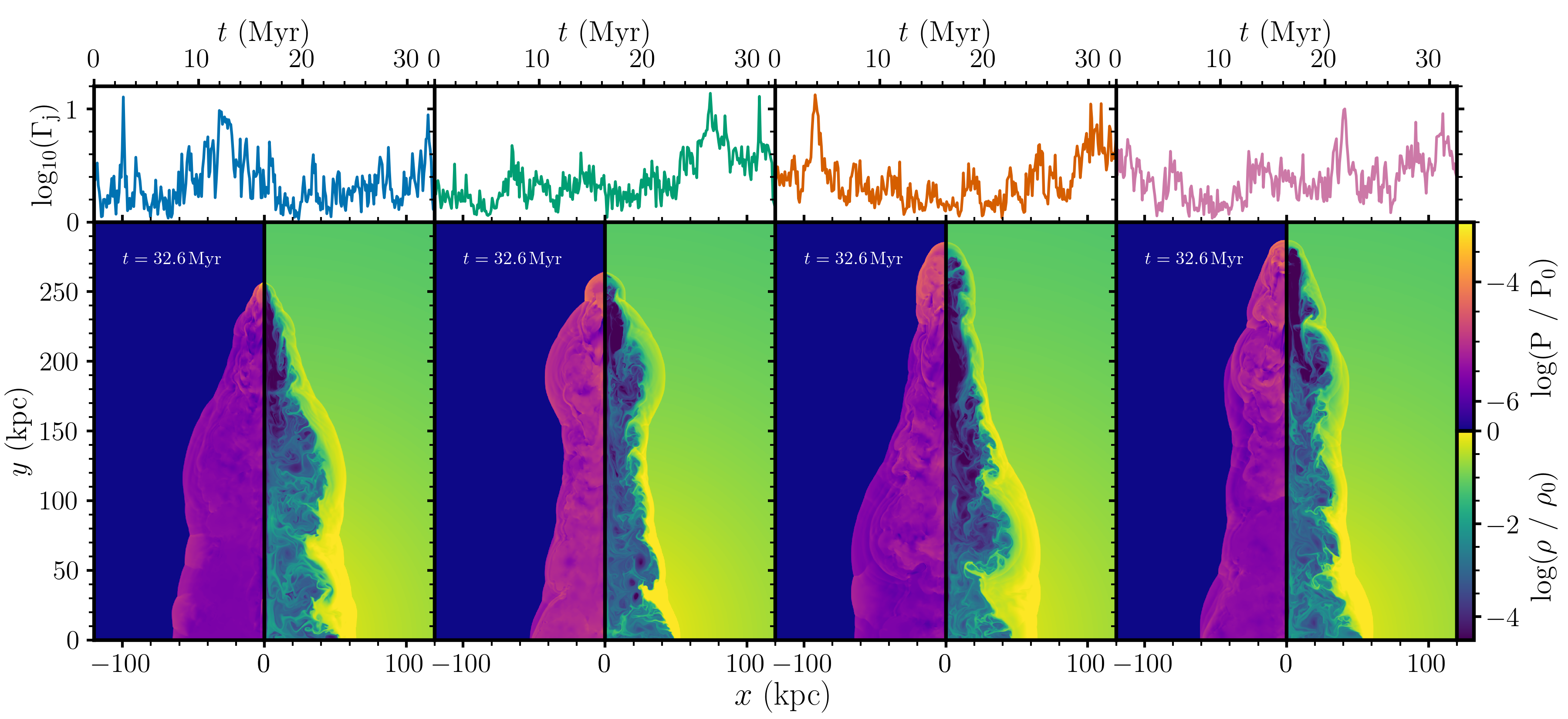}
    \caption{
    The effect of historical jet variability on the jet-cocoon system. The bottom panels show snapshots of (logarithmic) density and pressure at $t=32.6$\,Myr from each of the $\sigma=1.5$ simulations. The top panels show the corresponding jet power time series, with colours matching those in Fig.~\ref{fig:power_series}. The colour maps show the logarithm of $P/P_0$ and $\rho/\rho_0$ where $P_0$ and $\rho_0$ are the simulation units of pressure and density, respectively. In general, jets with earlier outbursts inflate wider jet bases, while recent outbursts create bulges or wider features near the jet head. An animated version of the figure is given in the supplementary material 
    (fig3\_animated.mp4). 
    }
    \label{fig:4panel_sigma1.5}
\end{figure*}

\begin{figure*}
    \centering
    \includegraphics[width=1.0\linewidth]{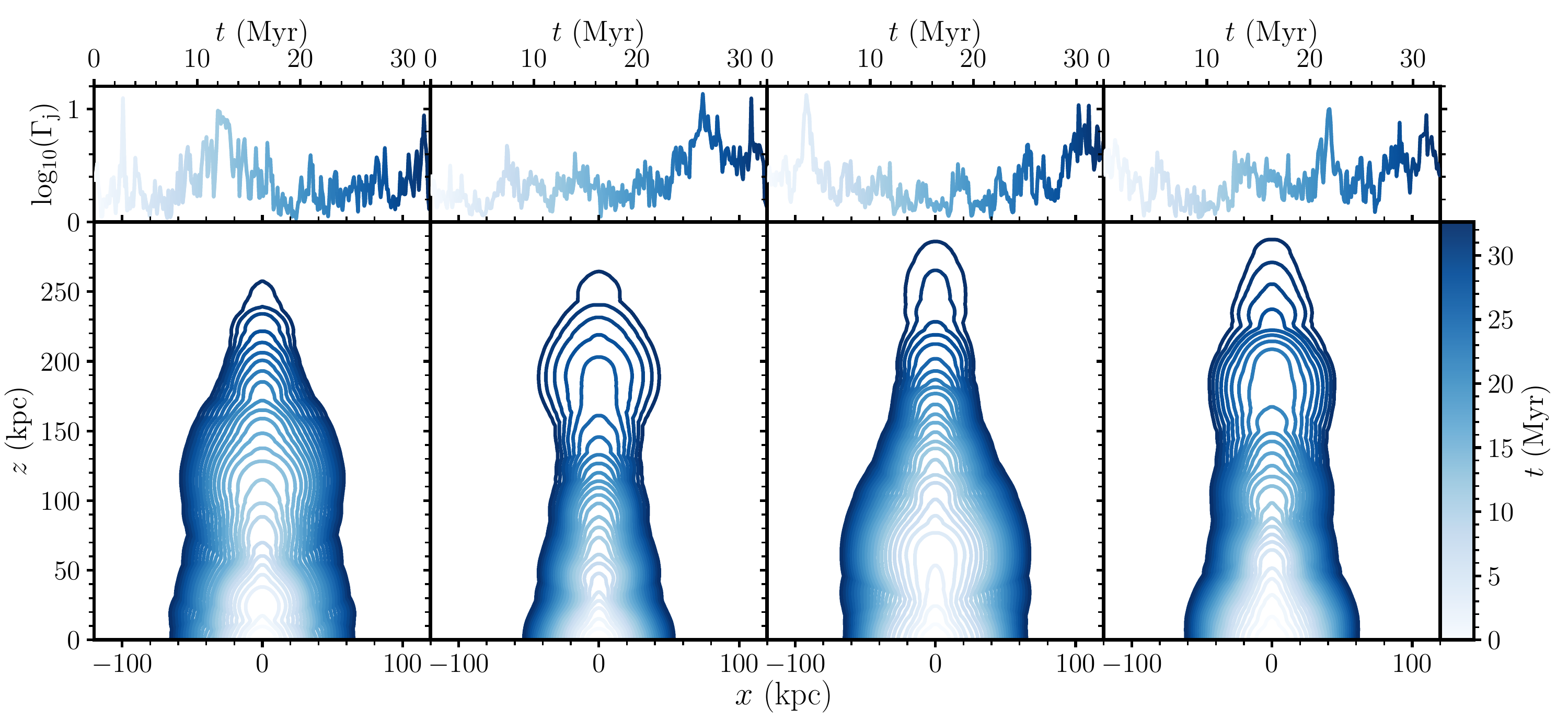}
    \caption{
    The impact of jet variability on bow shock morphology. The bottom panels show the outline of the bow shock at $1.31$\,Myr intervals from each of the $\sigma=1.5$ simulations, colour-coded by $t$. The top panels show the corresponding jet power time series, with the same colour-coding. Both the order of panels from left to right, and the final outline shown ($t=32.6$\,Myr), match Fig.~\ref{fig:4panel_sigma1.5}. 
    }
    \label{fig:4panel_outline}
\end{figure*}

\subsection{Overall behaviour and qualitative description}

To examine the overall behaviour of our jets, we first focus on the runs with the most dramatic variability $\sigma=1.5$. We show the logarithms of density and pressure at $t=32.6$\,Myr in Fig.~\ref{fig:4panel_sigma1.5} together with the jet power time series for each run. The general structure of the cocoon-jet system is similar to that produced in other simulations. The jet inflates a low density, high pressure cocoon and drives a bow shock into the ambient medium. Mixing at the cocoon-SAM interface proceeds via Kelvin-Helmholtz instabilities, increasing the density in the cocoon and creating a ragged, turbulent edge to the low density bubble. The jet deposits mechanical energy in a termination shock, establishing a pressure gradient between the jet head and cocoon, which drives strong backflows. Deep inside the cocoon, the plasma is transonic or subsonic, and the environment is characterised by gentle turbulence and vorticity. Prominent vortex rings can be seen in a few cases (for example in the second panel from the left). Overall, the general behaviour of the simulations is similar to the hydrodynamic simulations carried out to date that we described in the introduction.

Despite this consistent general behaviour, we can already see some of the impact of jet variability. Although the jet power time series have statistically similar properties in each of these four cases, the precise locations of peaks and troughs leave a clear imprint on the cocoon and bow shock morphology. Generally speaking, wider regions of the cocoon can be mapped to corresponding spikes or `outbursts' in the jet power or Lorentz factor. 
This behaviour was already seen to an extent in our semi-analytic work \citep{matthews_particle_2021}, and the phenomenon of bubbles linked to jet outbursts has been observed in simulations of intermittent jets \citep[e.g.][]{mendygral_mhd_2012,english_numerical_2019,yates_observability_2018}. To examine the shape of the bow shock more concretely, we show its location over time in Fig.~\ref{fig:4panel_outline}, at $1.31$\,Myr intervals. The panels match those in Fig.~\ref{fig:4panel_sigma1.5} and the colour-coding denotes the simulation time. The spacing of the lines in the vertical direction gives us a feel for the advance of the jet -- powerful outbursts lead to fast advance speeds, but these outbursts also leave a lasting imprint in the bow shock morphology. This is one of the first results of our work: the bow shock and, to a lesser extent, the cocoon, retain a memory of the jet outburst history, implying that radio galaxy morphologies could be used as a tool for understanding AGN fuelling and jet variability on $\gtrsim {\rm Myr}$ timescales. 

\begin{figure*}
    \centering
    \includegraphics[width=\linewidth]{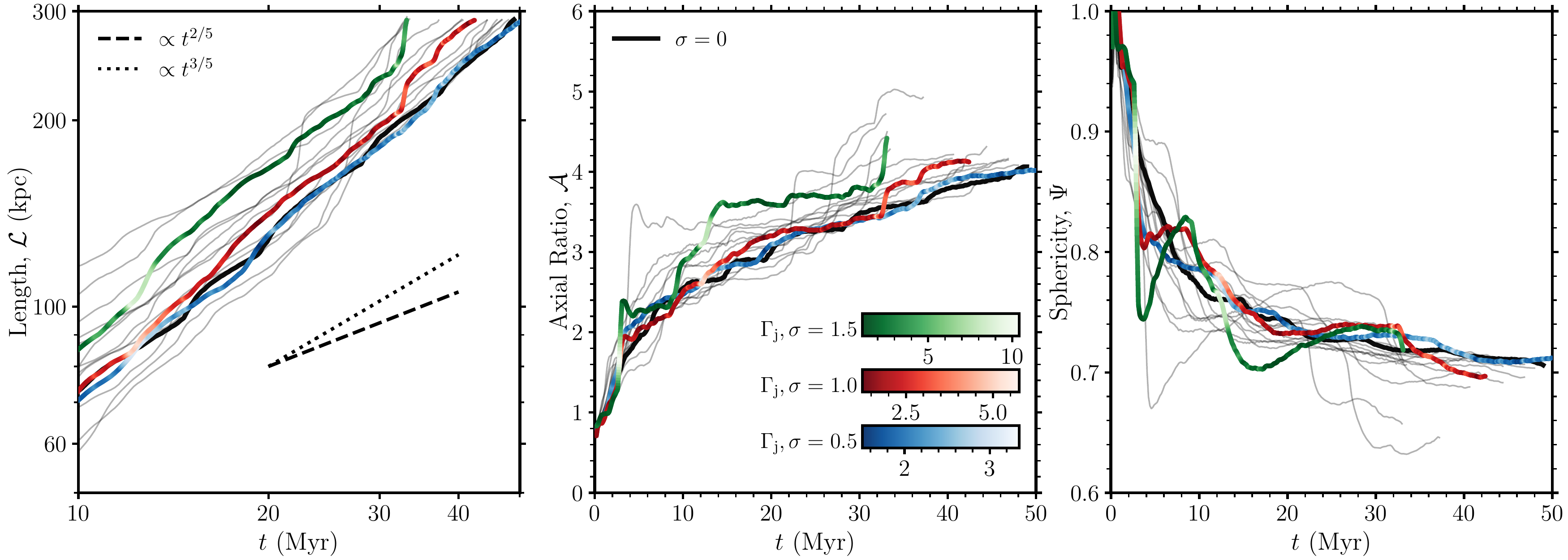}
    \caption{The impact of variability parameter $\sigma$ on jet and cocoon morphology. The panels show, left to right, the length of the jet, axial ratio of the cocoon and sphericity, as a function of time. All the simulations in the 2D grid are plotted as translucent grey lines, with the different values of $\sigma$ for one specific random number seed highlighted in colour. For each non-zero $\sigma$, the colourmap shows the bulk Lorentz factor of the jet at the jet inlet, $\gammajet$, with lighter colours corresponding to higher $\gammajet$.}
    \label{fig:oneseed}
\end{figure*}

\begin{figure*}
    \centering
    \includegraphics[width=\linewidth]{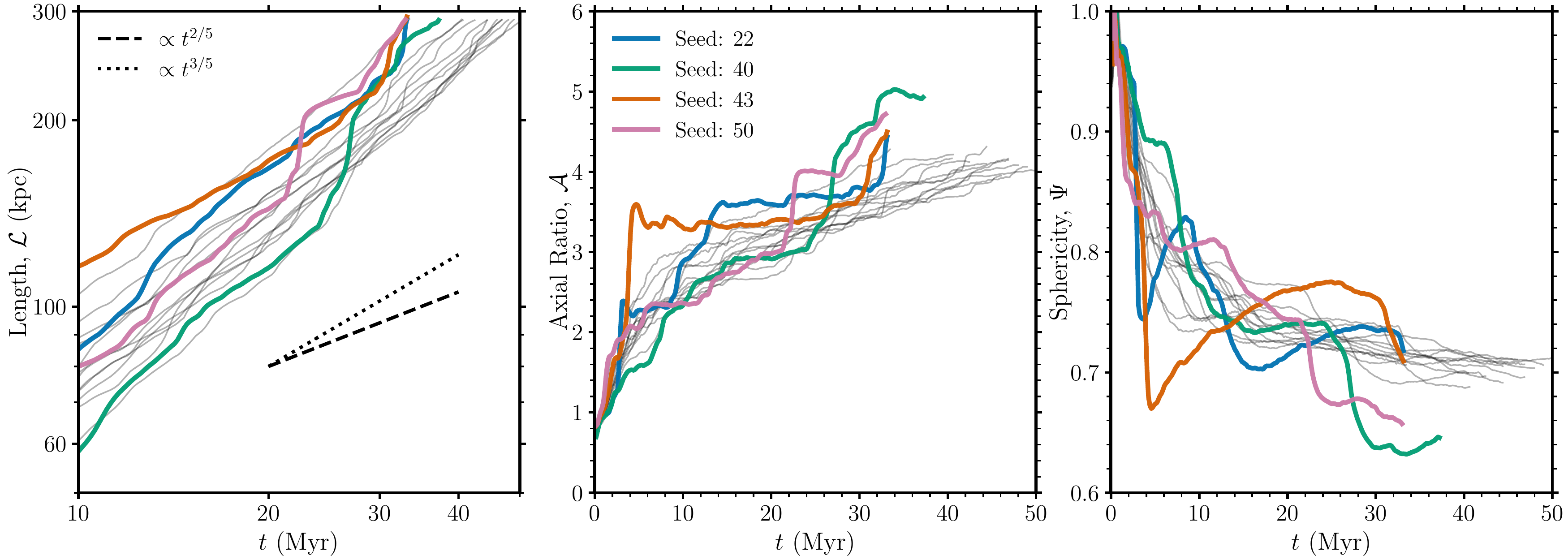}
    \caption{
    As Fig.~\ref{fig:oneseed}, but showing the impact of random number seed on jet and cocoon morphology. The panels show, left to right, the length of the jet, axial ratio of the cocoon and sphericity, as a function of time. All the simulations in the 2D grid are plotted as translucent grey lines, with the four runs with $\sigma=1.5$ highlighted in colour.
    }
    \label{fig:onesig}
\end{figure*}

\subsection{Measuring morphological evolution}
We make use of three different quantities to characterise the morphology of the bow shock: the jet length, ${\cal L}$, the axial ratio ${\cal A}$, and the sphericity, $\Psi$. The axial ratio is defined as ${\cal A}={\cal L}/{\cal W}$ where ${\cal W}$ is the width of the bow shock at the jet base. Sphericity is, as the name suggests, a measure of how spherical an object is and was introduced by \cite{1935JG.....43..250W} to quantize the shape of quartz crystals. Specifically, sphericity is the ratio of a 3D shape's surface area to the surface area of a sphere with the same volume as the shape. It is bound from above by $1$, with smaller values indicating less spherical shapes; for reference, a tetrahedron has $\Psi \approx 0.67$ and a hemisphere has $\Psi \approx 0.79$. We calculate sphericity by identifying the bow shock boundary and evaluating the  volume ${\cal V}$ and surface area ${\cal S}$,
\begin{equation}
    \centering
    \Psi = \frac{\pi^{\frac{1}{3}}\left(6 {\cal V}\right)^{\frac{2}{3}}}{{\cal S}},
\end{equation}
where the volume and surface area are evaluated over both hemispheres assuming a perfectly reflected shape about $z=0$. 

In Figs.~\ref{fig:oneseed} and \ref{fig:onesig}, we show ${\cal L}$, ${\cal A}$ and $\Psi$ from left to right as a function of time for our simulation grid. Every member of the simulation grid with $\sigma>0$ is shown in each panel, but we highlight runs with a fixed RN seed, $i$, and varied $\sigma$ in Fig.~\ref{fig:oneseed} and we highlight runs with a fixed $\sigma=1.5$ and varied $i$ in Fig.~\ref{fig:onesig}. In Fig.~\ref{fig:oneseed} we colour-code the highlighted lines by the value of $\log_{10} (\gammajet)$ to show the periods of high and low activity. High power episodes correspond clearly to steeper gradients in ${\cal L}$ (faster advance speed $v_{\rm adv} = d{\cal L}/dt$), whereas in quiescent periods the gradient is shallower and in some cases approaches a Sedov-Taylor like scaling of ${\cal L}\propto t^{2/5}$. High power episodes also lead to sharp increases in the axial ratio ${\cal A}$ as the jet advances rapidly with only modest lateral expansion. Similarly, the sphericity, $\Psi$, decreases quickly in periods of high activity and relaxes or increases in low power periods. The overall dynamics of the jet are similar to the swimming action of a jellyfish \citep[e.g.][]{rakow} -- in periods of high activity the jet head narrows and the jet advances forwards quickly; then, if/when the jet power drops, the cocoon expands laterally, gradually becoming more spherical and advancing more slowly. This behaviour imprints itself on the morphology of the jet as a whole, as can be seen in Fig.~\ref{fig:4panel_sigma1.5}, where, as mentioned above, the bow shock and the cocoon both retain some memory of the outburst history. Wide structures at the base are present in simulations with an early outburst, whereas wide structures near the jet head correspond to recent periods of powerful activity. 

The response of the jet to power variations -- and the relative dominance of forward versus lateral motion -- can be readily understood in terms of the relative importance of hot-spot pressure and average cocoon pressure. The former is determined by the instantaneous ram pressure of the jet at the termination shock, whereas the latter is an integrated quantity set by the total energy injected into the cocoon divided by the volume, less the $P~dV$ work done on the surroundings. Thus, in periods of fast advance we expect the largest pressure difference between the hotspot and cocoon. This pressure gradient is also responsible for the creation of backflow from the jet head into the cocoon as discussed in the next subsection. 

The possible parallels between Sedov-Taylor blast waves and intermittent jets are worth discussing. Truly intermittent or dormant sources can show a quasi-spherical Sedov-Taylor-like expansion if they are over-pressured with respect to their surroundings \citep{reynolds_hydrodynamics_2002,yates_observability_2018,english_numerical_2019}; then, once pressure equilibrium with the surroundings is reached, a transition to a buoyant rising bubble occurs \citep{reynolds_hydrodynamics_2002}. The cocoons in our simulations are always over-pressured, so we never observe this buoyant phase. A true Sedov-Taylor phase is also rare, as can be inferred from the left panels of Figs.~\ref{fig:oneseed} and \ref{fig:onesig}; there the canonical ${\cal L} \propto t^{2/5}$ Sedov-Taylor scaling is plotted in a dotted line. Although the length evolution sometimes starts to approach this scaling, in the majority of cases the jet advance is much faster than $t^{2/5}$. A notable exception can be found in the seed $50$, $\sigma=1.5$ simulation (red line, Fig.~\ref{fig:onesig}, where, at around $24$\,Myr a sudden injection of energy is followed by a Sedov-Taylor phase that approaches the self-similar slope. The relative scarcity of Sedov-Taylor phases can be partially attributed to our choice of jet power PSD; a set of more discrete outbursts or a redder PSD slope would lead to behaviour resembling a series of short impulses and allow Sedov phases to occur. In supernova remnants, the approach towards self-similarity can be slow \citep[e.g.][]{bell_cosmic_2015}. In our simulations, the pink noise spectrum adopted means that jet power spikes occur quite frequently and the system does not have time to relax towards the self-similar solution. This behaviour highlights the general link between morphology and the power spectrum of AGN accretion, a link which we discuss further in section~\ref{sec:discuss_fuel}.  

\subsection{Dynamics: backflow, shocks and turbulence}
\label{sec:2d_dynamics}

\begin{figure*}
    \centering
    \includegraphics[width=\linewidth]{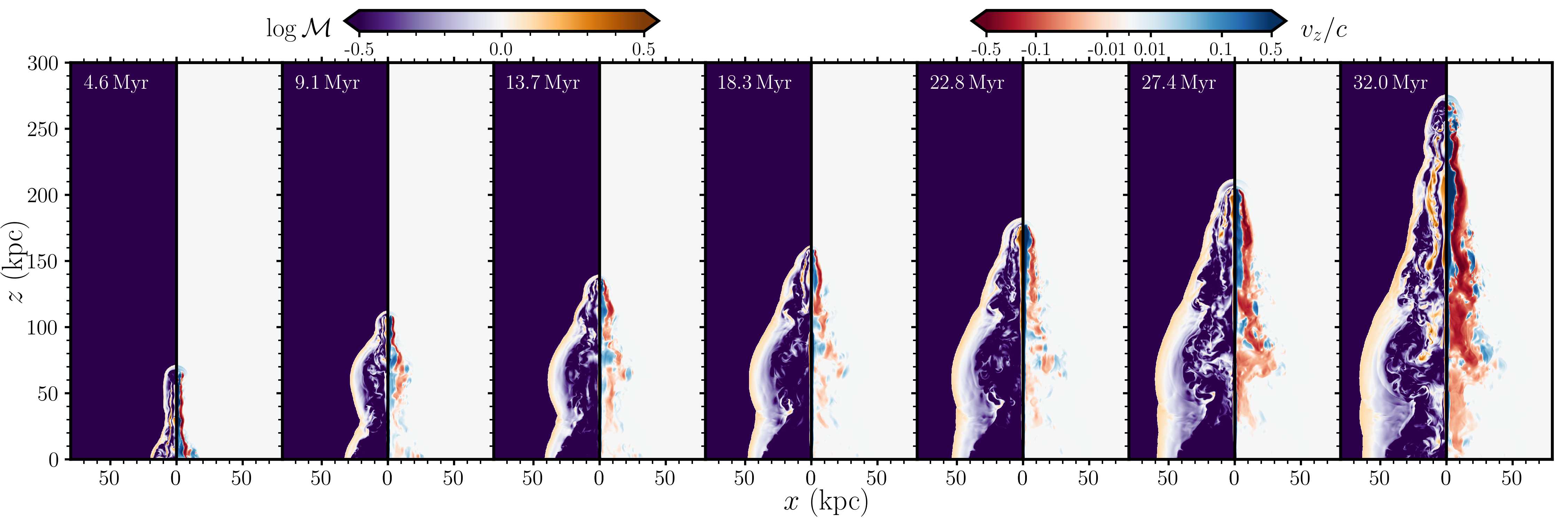}
    \caption{
    Simulation snapshots, from left to right, at $4.56$\,Myr intervals for $i=43$, $\sigma=1.5$. Each pair of colourmaps shows the logarithm of the Mach number, $\log_{10} {\cal M}$ (left), and the vertical velocity component $v_z/c$ (right) on a symmetric logarithmic colour scale. 
    % An animated version of this figure (fig7\_animated.mp4), and 
    The corresponding figures for $i=40$ and $\sigma=0.5$ are supplied in the supplementary material. 
    }
    \label{fig:ColorMeshTime}
\end{figure*}

\begin{figure}
    \centering
    \includegraphics[width=\linewidth]{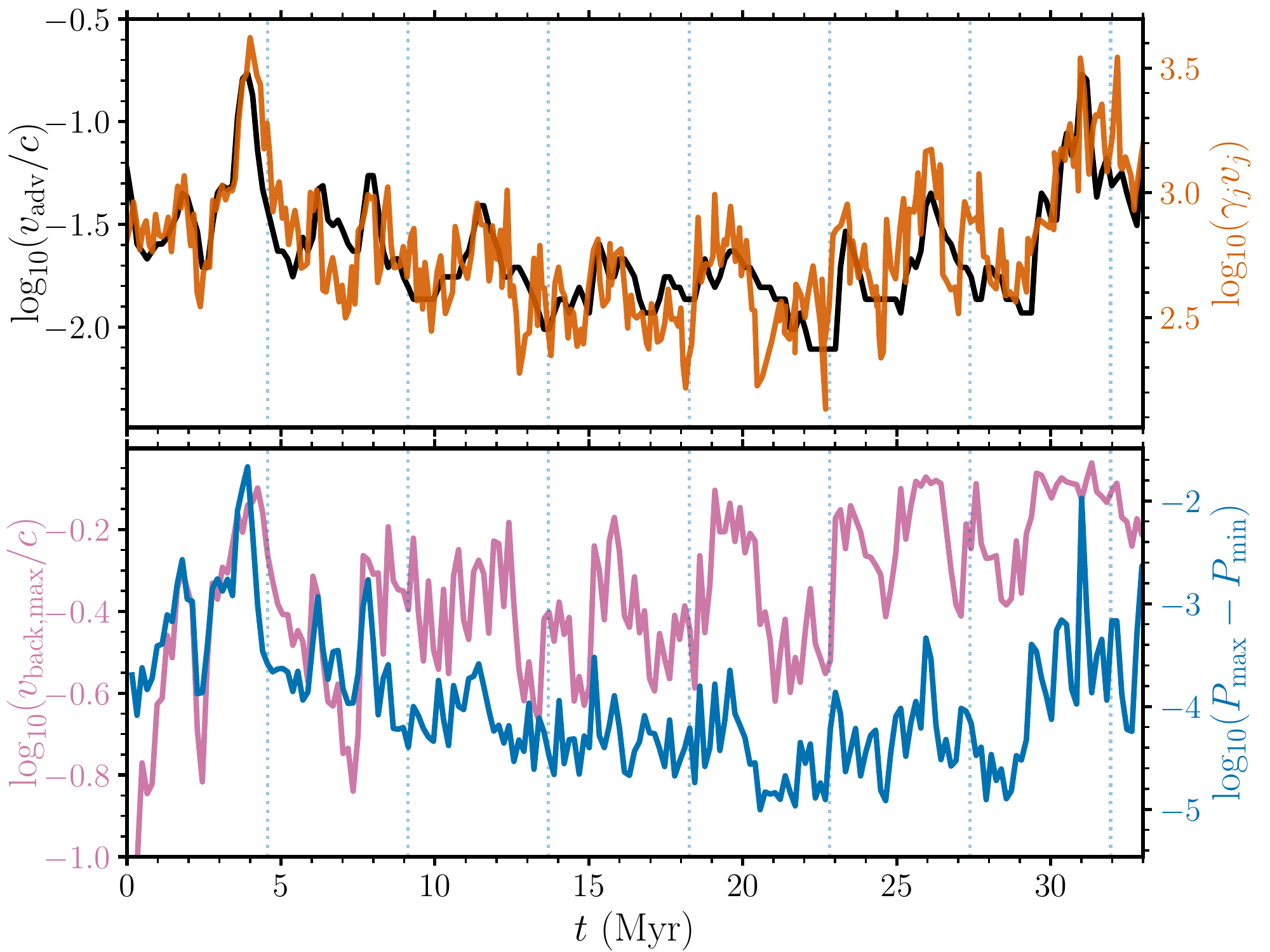}
    \caption{An illustration of how jet advance and cocoon dynamics respond to input power variations for $i=43$, $\sigma=1.5$. Top: Comparison of true shock front advance (black, left axis) to relativistic momentum at jet inlet. The jet inlet values are delayed by a characteristic jet travel time, crudely estimated as ${\cal L} / v_{\rm j}$, the ratio of instantaneous jet length and jet inlet velocity. Bottom: Comparison of the maximum backflow velocity (red, left axis) and range of cocoon pressures (blue, right axis), over the same time period as the top panel. Periods of strong backflow occur when pressure gradients are large, which in turn occur when the jet power is in a high state. In both panels the dotted vertical lines mark $4.57$\,Myr intervals, for which snapshots are plotted in Fig.~\ref{fig:ColorMeshTime}.
    }
    \label{fig:advance_with_backflow}
\end{figure}

Jets can create strong, supersonic backflows -- flows away from the jet head towards the jet launch point -- with velocities an appreciable fraction of $c$ \citep{reynolds_hydrodynamics_2002,mignone_high-resolution_2010,cielo_3d_2014,matthews_cosmic_2019}. Backflow is an important process in radio galaxy cocoons; for example, it deposits mechanical energy in the lobes and creates turbulence and vorticity \citep[e.g.][]{falle_self-similar_1991}, transports radiating particles \citep{turner_raise_2018}, and can produce shocks and accelerate particles \citep{matthews_cosmic_2019,bell_cosmic_2019,mukherjee_new_2020}. Backflows may also be important for explaining complex morphologies; for example, they are a credible explanation for `X-shaped' radio galaxies \citep{leahy_bridges_1984,hodges-kluck_chandra_2010,hodges-kluck_hydrodynamic_2011,cotton_hydrodynamical_2020}.

Our simulations do produce fast, supersonic backflows, as expected, which can be seen in the upper panels of Fig.~\ref{fig:ColorMeshTime}, showing vertical velocity, $v_z$, and Mach number, ${\cal M}$.  Characteristically we see a backflow extending a significant distance, between $10$ and $50\%$ of the jet length, back towards the jet inlet. This backflow is at least partly supersonic and the phenomenon of `vortex shedding' is prevalent. Indeed, the general behaviour is a gradual transition from a focused supersonic beam to a subsonic or transonic turbulent medium made up of vortex rings and eddies. Despite this general behaviour, there is clear diversity in the detailed dynamics of the backflow and the scale length over which the backflow loses its identity. 

To examine quantitatively how the properties of the cocoon backflow change over time, we define the maximum backflow velocity as 
\begin{equation}
v_{\rm back} = {\rm max} (-v_{z,c})
\label{eq:backflow}
\end{equation}
and we define the pressure range in the cocoon as 
\begin{equation}
\Delta P_c= {\rm max} (P_c) - {\rm min} (P_c).
\label{eq:deltaP}
\end{equation}
In both these equations, the $c$ subscript denotes a cocoon value, and corresponds to having a jet tracer value that satisfies $0.95>C_{\rm j}>10^{-4}$. The adopted threshold of $10^{-4}$ is somewhat arbitrary, since the cocoon-SAM interface is a complex mixing layer, and was chosen by eye so as to roughly delineate the low density cocoon. We experimented by changing this threshold and found our results are not sensitive to its precise value as long as it is in the approximate range $10^{-3}$ to $10^{-6}$. The evolution over time of $v_{\rm back}$ and $\Delta P_c$ is shown on twin axes in the bottom panel Fig.~\ref{fig:advance_with_backflow}.
The pressure difference is strongly correlated with the advance speed and input jet velocity. 
At early times, the backflow speed is also correlated well with the pressure difference $\Delta P_c$, but at late times this correlation starts to break down. This apparent decoupling is partly a result of our adopted definition of $\Delta P_c$, since the minimum pressure can be deep in the cocoon and offer limited information on the pressure gradient governing backflow near the jet head; however, the phenomenon of fast, persistent backflow at late times is more generic. 

Our results show that fast backflows are preferentially produced during periods of powerful jet activity and fast advance -- as would be expected superficially from Fig.~\ref{fig:ColorMeshTime}). However, in addition, we have shown that the backflow dynamics are likely to depend sensitively on the detailed shock and instability physics relating to the complex backflow-jet interaction, rather than just the macroscopic pressure gradient. \cite{mizuta_hysteresis_2010} have explored the backflow-jet interaction in detail. They found that when the jet advance speed is slow compared to the hotspot sound speed (as is generally the case for the light jets modelled here), a bent backflow results and oblique shocks are formed near the jet head. This complex behaviour could be enriched further by variability and may explain some observed features in radio galaxy images and spectral index maps (see section~\ref{sec:discuss_obs}). 

\begin{figure*}
    \centering
    \includegraphics[width=0.8\linewidth]{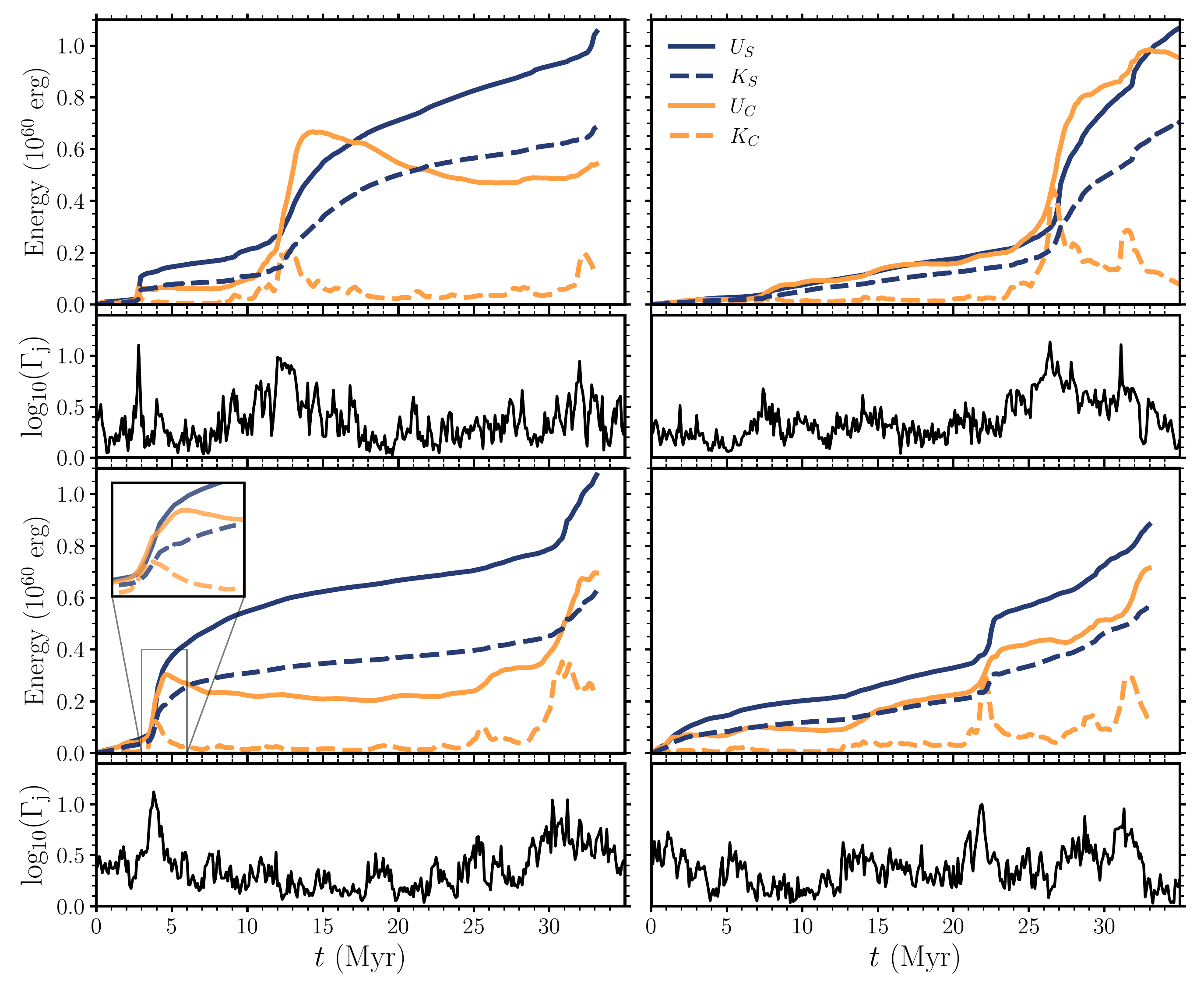}
    \caption{Partitioning of energy by region and type (internal or kinetic) for the most variable ($\sigma$ = 1.5) members of the 2D grid, separated between subplots by seed. The central plots show the growth of the various energy reservoirs energies over time, split between the internal energy of the SAM ($U_s$), the kinetic energy of the SAM ($K_s$), the cocoon internal energy ($U_c$), and the cocoon kinetic energy ($K_c$). The cocoon includes the jet itself and so $K_c$ is usually dominated by the jet kinetic energy. The outer panels show the time series of $\log_{10} \gammajet$ in each case. Comparison of the outer panels to the relevant central panel shows the response of the jet, cocoon and SAM material to the input power variations. In the lower left panel, an inset figure shows a zoom in of the response of the various energy reservoirs to a jet impulse, showing how the cocoon responds quickly whereas the SAM responds more slowly.}
    \label{fig:partitioning}
\end{figure*}

\subsection{Energetics}
To consider how energy is partitioned in the system, we calculate the lab frame kinetic ($K$) and internal ($U$) energies in the cocoon and shocked ambient medium (denoted with subscripts $c$ and $s$, respectively). Kinetic energies are calculated as 
\begin{equation}
    K = \sum_n \Gamma_n \left( \Gamma_n - 1 \right) \rho_n c^2 \Delta {\cal V}_n \,
\end{equation}
where $\Delta {\cal V}$ is the cell volume and $n$ is a cell index. 
The total internal energy $U$ is calculated from 
\begin{equation}
    U = \sum_n \Gamma_n^2 (\rho e)_n \Delta {\cal V}_n ,
\label{eq:u}
\end{equation}
where the rest frame internal energy density in each cell, $\rho e$, is calculated by inverting the Taub-Mathews equation of state \citep{taub_relativistic_1948,mignone_equation_2007}, giving the equation
\begin{equation}
    \rho e = \frac{1}{2} \left(3P -2\rho c^2 + \sqrt{\left(3P - 2\rho c^2\right)^2 + 12P\rho c^2}\right).
\end{equation}
For both $U$ and $K$, the sum is over all cells within the the relevant region (cocoon, jet, SAM; see section~\ref{sec:regions}), as indexed by the subscript $n$. Note that $K+U$ is not actually a conserved quantity for $\Gamma>1$ (as can be inferred from equation~\ref{eq:conserved}); however, for our purposes most of the internal energy is contained within non-relativistic plasma and so $K+U$ is fairly close to the sum of the total energy $E$. In addition, the results are almost identical if the $\Gamma^2$ term is omitted from equation~\ref{eq:u}. 

In Fig.~\ref{fig:partitioning}, we show the evolution of $K_c$, $U_c$, $K_s$ and $U_s$ in the four $\sigma=1.5$ simulations, with the corresponding Lorentz factor time series also shown. The differing responses of the energy reservoirs to the jet power variations are apparent from these plots. The kinetic energy of the cocoon (which is generally dominated by the kinetic energy of the jet) is essentially a smoothed version of the input power time series, as we would expect. This energy is transferred into cocoon internal energy, $U_c$, which responds with a slight delay, and more slowly, compared to $K_c$. Although the general trend is for $U_c$ to increase over time, there are occasions where the value of $U_c$ decreases, as the amount of work done via $P dV$ expansion exceeds the rate of energy input from the jet. The over-pressured cocoon is responsible for driving the bow shock into the ambient medium, and thus the energies in the shocked medium, $U_s$ and $K_s$, gradually increase over time with a slower response still. 

In simulations of steady jets, the ratio of the internal energies and kinetic energies between the cocoon and shocked medium ($U_c/U_s$ and $K_c/K_s$) are generally close to unity and roughly constant with time \citep{hardcastle_numerical_2013}. Our simulations show that variability can disrupt this close coupling of cocoon and shocked medium energies. Shortly after jet outbursts, the internal energy of the cocoon increases rapidly and can dominate over $U_s$. The `thermalisation' of jet kinetic energy into cocoon internal energy, i.e. $K_c \to U_s$, takes place on roughly a jet crossing time ${\cal L}/v_{\rm j} \approx {\cal L} / c$. On long timescales, $U_s$ is generally dominant. The cocoon energetics are generally more sensitive to recent activity (over the last few Myrs), whereas the shocked medium acts as a {\em calorimeter} and tracks the overall energy input. These results have implications for observational estimates of jet power and scaling relationships between jet power and radio luminosity, which we explore further in section~\ref{sec:jet_power}. 

% \clearpage
\section{Results from a fiducial 3D simulation} 
\label{sec:3d_results}
We now present results from a single `fiducial' 3D simulation, with $i=40$ and $\sigma=1.5$. 

\subsection{Comparison with 2D results}
We begin by comparing our 3D results with the appropriate 2D simulation with the same $\sigma$ and RNG seed. In the 3D run, we encountered some numerical problems in which unphysical (negative density or pressure) states were occasionally produced in a few cells. These problems can be produced in the presence of strong gradients \citep[see e.g. Appendix B of ][]{mignone2012}, and so in some sense it is not surprising that introducing variability in the jet Lorentz factor -- which is sometimes quite dramatic -- creates numerical issues. We attempted to fix this with the inbuilt \pluto\ procedures that re-solve the Riemann problem in and around problem cells with a more dissipative scheme, and while this improved the stability somewhat, we still found occasional problem cells. To circumvent these issues, we tried very slightly smoothing the Lorentz factor time series, and found this did allow our simulation run to completion without problems within the $200$~kpc long domain. We smoothed the time series with a Savitzky-Golay filter with a window length of 7 data points ($0.7$~Myr) using a 3rd order polynomial. Our approach here is slightly crude, but it allows us to produce a simulation with the desired overall qualities and only subtly different quantitative outputs, leaving the overall science results and astrophysical implications unaffected. 

\begin{figure*}
    \centering
    \includegraphics[width=0.49\linewidth]{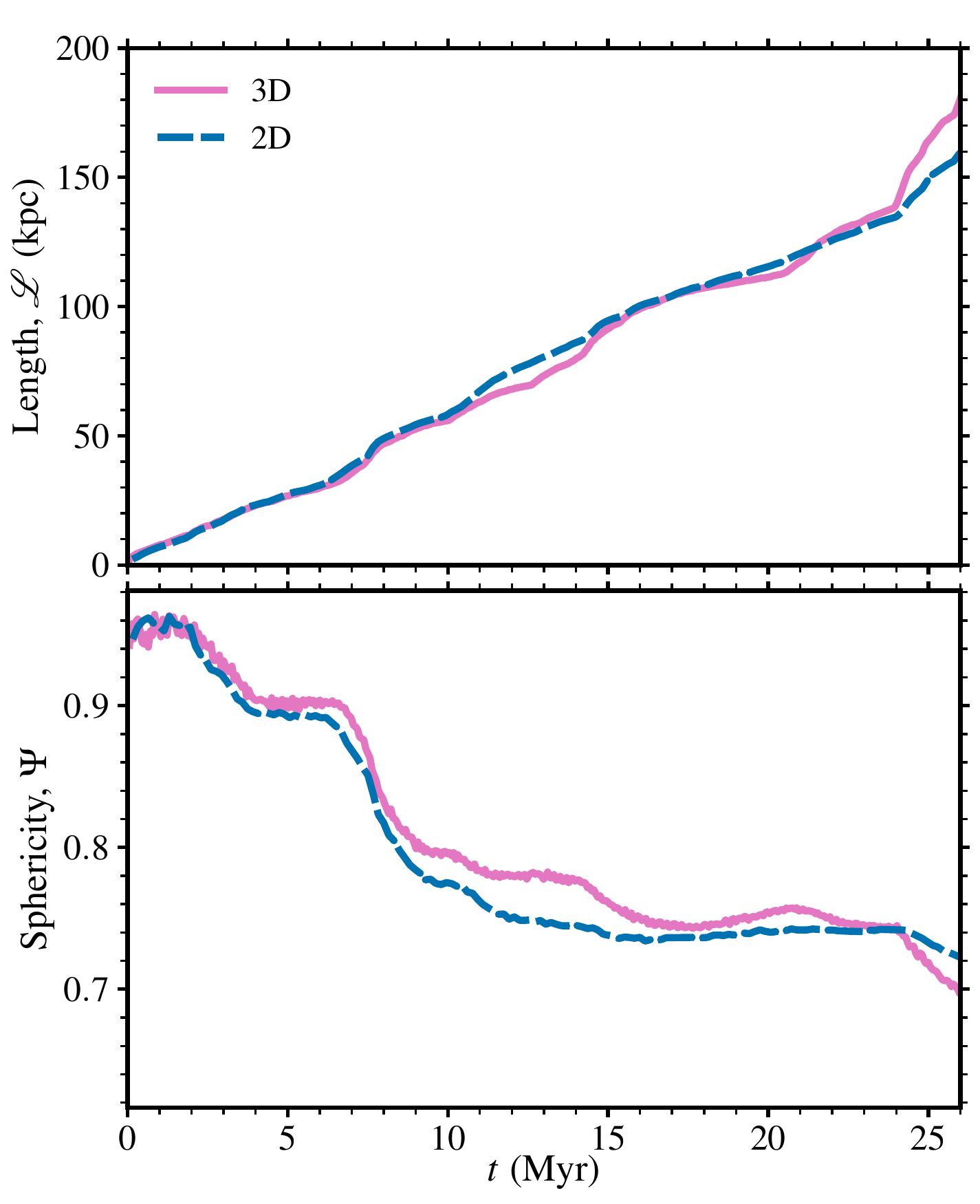}
    \includegraphics[width=0.49\linewidth]{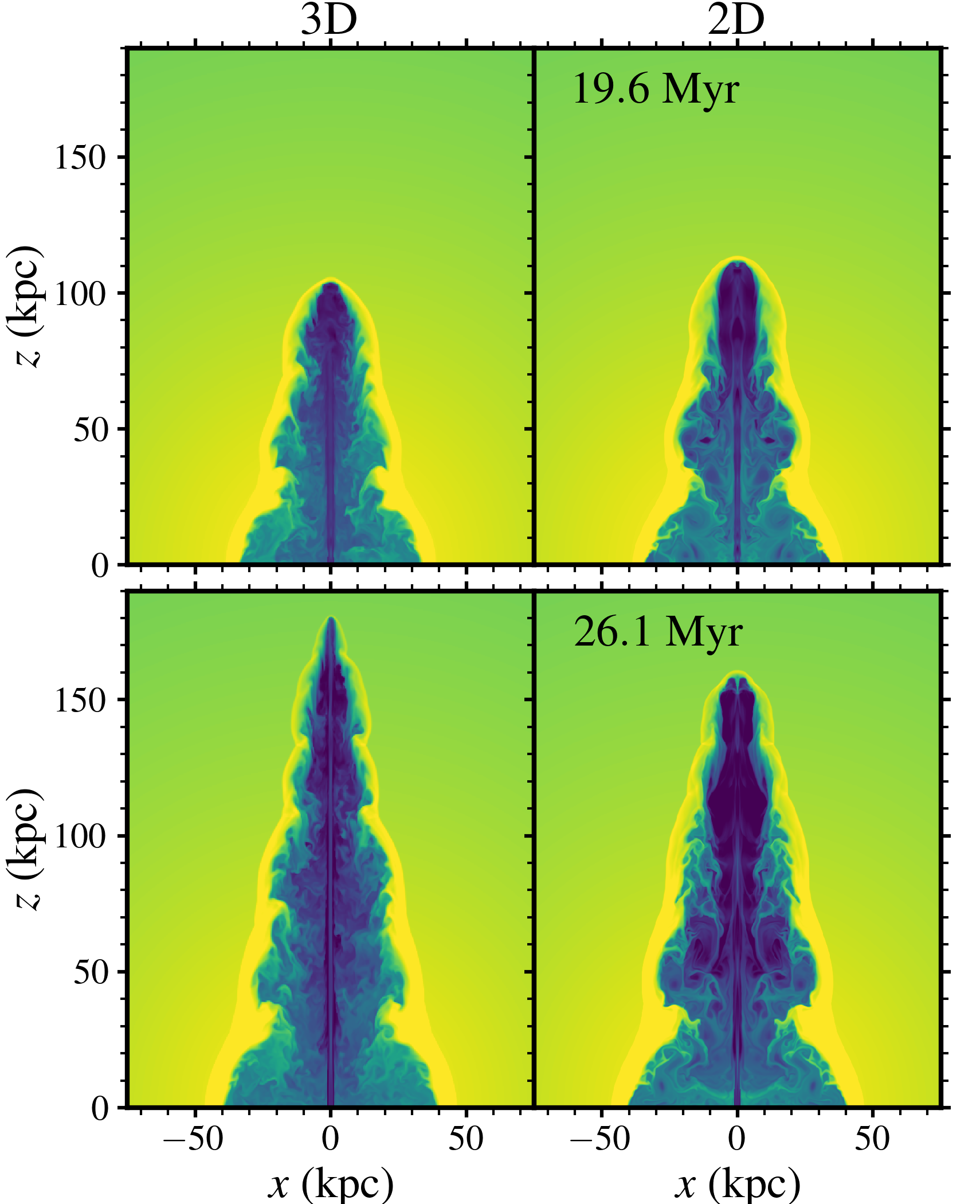}
    \caption{{\sl Left:} Morphological evolution for 3D simulation as compared to 2D simulation with same input jet power time series. The length, ${\cal L}$, and sphericity, $\Psi$, are plotted over time, showing that, in general, the 2D and 3D simulations show similar behaviour and agree well. {\sl Right:} The logarithm of density, on the same colour scale as Fig.~\ref{fig:4panel_sigma1.5}, in the 2D and 3D simulations, at $19.58$~Myr (top) and $26.11$~Myr (bottom). At early times the shapes of the cocoon and bow shock are extremely similar, but in the later snapshot the 3D simulation starts to advance significantly more quickly. 
    }
    \label{fig:comparison}
\end{figure*}

In the left-hand panel of Fig.~\ref{fig:comparison}, we show a comparison of the jet length and sphericity measured from the 3D simulation as compared to the 2D simulation. We focus on sphericity here rather than the more familiar axial ratio, because it does not require choices about where to evaluate the jet width and provides a simple metric to describe complex morphologies. The length and sphericity evolution of the 3D simulation closely tracks the 2D simulation, in general, although the 3D simulation is slightly more variable in terms of its advance. At $t\approx25~{\rm Myr}$, correlating with an increase in $\gammajet$ the 3D simulation starts to advance more quickly than the 2D simulation (leading also to a corresponding drop in sphericity). Differences in advance speeds between 2D and 3D simulations are expected. The relaxation of the axisymmetric assumption means that 3D effects within the beam such as helical motion or `wobbling' can act to speed up jet head propagation \citep{aloy_high-resolution_1999,perucho_long-term_2019}, so a difference in jet propagation at some level is not surprising. It is also possible that the slightly different resolutions in the two simulations ($0.3125$\,kpc in 3D versus $0.195$\,kpc in 2D), could lead to some differences in advance speed as has been found in other numerical studies \citep{krause2001,donohoe2016}. The overall qualitative agreement of the length and sphericity evolution here suggests many of our main science conclusions from the 2D simulations can be generalised to a more realistic 3D system. 

To test this statement further, we can also examine the morphology and dynamics of the jet-cocoon system to see if the 2D phenomenology holds in 3D. In the right-hand panel of Fig.~\ref{fig:comparison}, we show the logarithmic density from two snapshots of the simulations, with the 2D and 3D versions plotted side-by-side. In the earlier snapshot, the shape of the bow shock and structure of the cocoon-SAM interface are extremely similar. Turbulent mixing is proceeding in a similar way in both cases and the ragged cocoon edge bears an obvious cosmetic similarity. The later snapshot corresponds to the period when the 3D simulation has started to outstrip the 2D one, and so in this case the striking similarity is no more. The 3D jet is significantly longer and has a different structure near the jet head. 
However, we take these results as evidence that the {\em qualitative} behaviour, and the main scientific results from section~\ref{sec:2d_results}, are indeed applicable to 3D.

\subsection{3D Dynamics}
\label{sec:3d_dynamics}

To examine the dynamics of the 3D simulation, we show the vertical velocity component ($v_z$) and logarithmic of the Mach number, $\log {\cal M}$ in Fig.~\ref{fig:backflow_3d}. The figure is a rough analogue to Fig~\ref{fig:ColorMeshTime}. The plots show slices through the computational domain at $y=0$, with timestamps from left to right from $10.4$~Myr to $26.1$~Myr at $2.61$~Myr intervals. By moving to a 3D cartesian grid with imposed density pertubations, the azimuthal symmetry is broken and the simulation has an asymmetric backflow. In addition, the jet is able to wobble and undergo helical motions. Discussions of 3D backflow dynamics and the breaking of aximuthal symmetry are given by other authors \citep[e.g.][]{matthews_ultrahigh_2019}, so here we instead focus mostly on the aspects of the simulations specific to this setup and induced by jet variability. 

The results shown in Fig.~\ref{fig:backflow_3d} are once again qualitatively rather similar to those from the 2D simulation. Fast, supersonic backflow is ubiquitous in time, but there is diversity in the detailed behaviour. In high-power periods (as in the $15.7$ and $26.1$~Myr panels), the backflow can extend in a fairly coherent manner for nearly $100~{\rm kpc}$, with transonic turbulence being induced deep into the cocoon. The (qualitative) vorticity of the flow is different in 2D and 3D as expected since any vortex rings are not constrained to move only in two dimensions and are more complex in shape \citep[see e.g.][]{melander_computational_1987}. \cite{falle_self-similar_1991} gives a thorough discussion of vortex shedding in a steady jet simulation, including its impact on jet propagation and self-similarity. \cite{falle_self-similar_1991} also estimates a vortex shedding timescale, finding the process should be most important earlier in the jet evolution. We instead find that, in this specific simulation, a degree of vortex shedding and turbulence appears to be prevalent late into the jet evolution due to the increase in power at  $\approx24$~Myr. Although the generation of vorticity and turbulence is not the primary focus of our work, our results imply that jet variability has an interesting impact on the induced turbulence and vorticity. The fact that this turbulence also interacts with the jet itself creates the intriguing possibility that any `feedback' here is emphasized during high-power episodes. 

There are also a few phenomena specific to variable 3D jets which are worth discussing. The jet is no longer restricted to axisymmetric motion, which means that helical motion can occur and instabilities can develop. These 3D hydrodynamic effects have been studied by various authors \citep[e.g.][]{aloy_high-resolution_1999,hughes_three-dimensional_2002,perucho_stability_2010} and can affect the stability of the jet beam and advance speed of the jet-cocoon system. The additional factor introduced here is that jets are more likely to be stable during high-power episodes. In some low states, for example in the $18.3$~Myr panel of Fig.~\ref{fig:backflow_3d}, the jet is disrupted, the backflow is weak or non-existent, and the jet fails to reach the end of the cocoon. This shows that jet variability can lead to intermittent power supply to the cocoon head, leading in turn to intermittent hot spots and an advance speed that is decoupled from the instantaneous jet power. We refer to these episodes as `jet discontinuities' and note their potential importance for the FRI/FRII dichotomy and hotspot prevalence in radio galaxies (see Section~\ref{sec:discuss_obs}). A simplified interpretation would be that for a jet to propagate undisturbed to the end of the cocoon its ram pressure must significantly exceed the thermal and non-thermal pressure in the cocoon. In reality, the disruption physics is more complex and depends on the growth-rate of helical modes within the jet beam as well as the details of the jet-cocoon interface. With strong magnetic fields, the situation is similar but dictated instead by the growth of the magnetic kink instability, which is not captured in our purely hydrodynamic simulations (see section~\ref{sec:limitations}). 

\begin{figure}
    \centering
    \includegraphics[width=\linewidth]{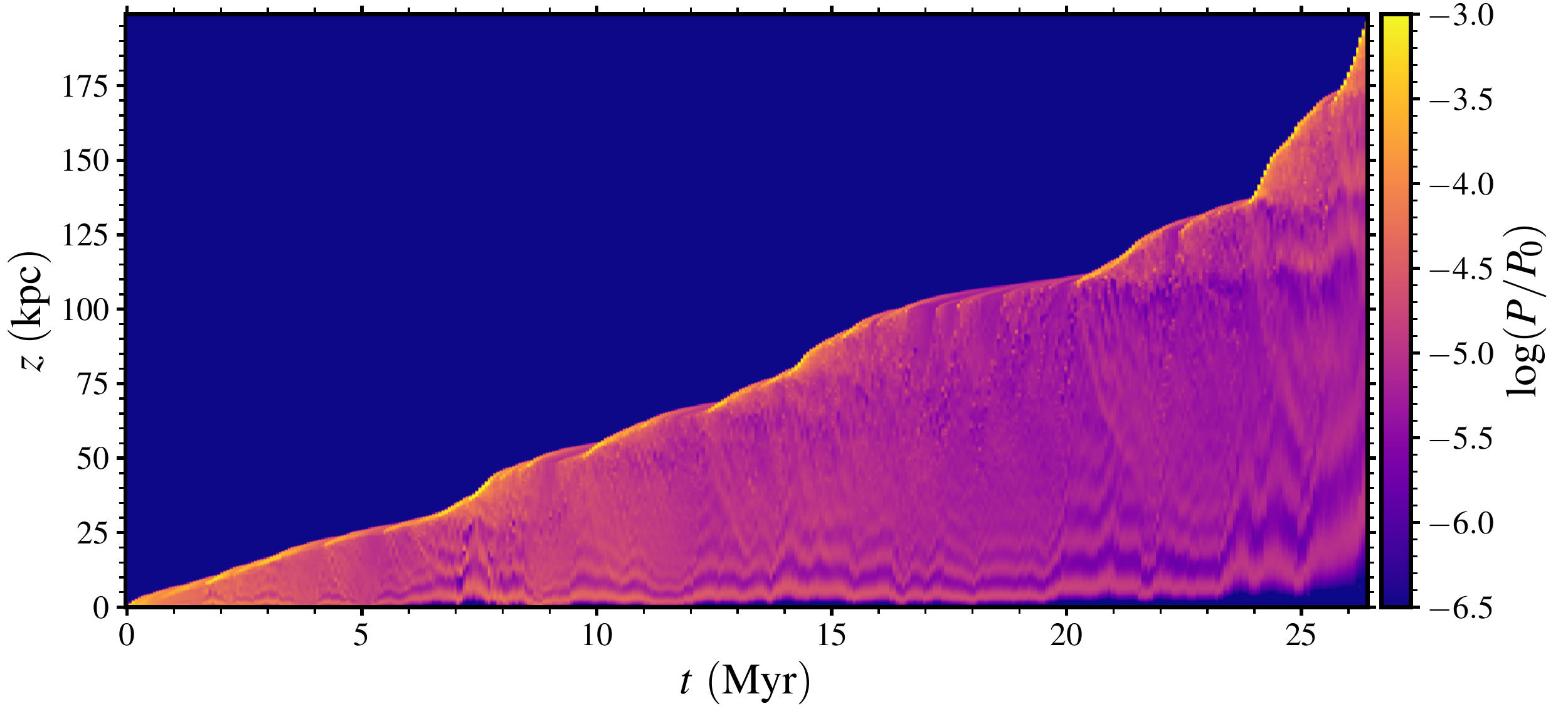}
    \caption{Space-time plot of the logarithm of the pressure along the 3D simulation jet column. We plot, as a colour map, the pressure profile along the central jet column at $0.065$~Myr intervals throughout the 3D simulation. The purpose is to visualise the time evolution of the shock structures and pressure changes along the jet beam.
    }
    \label{fig:spacetime}
\end{figure}

\begin{figure*}
    \centering
    \includegraphics[width=1.0\linewidth]{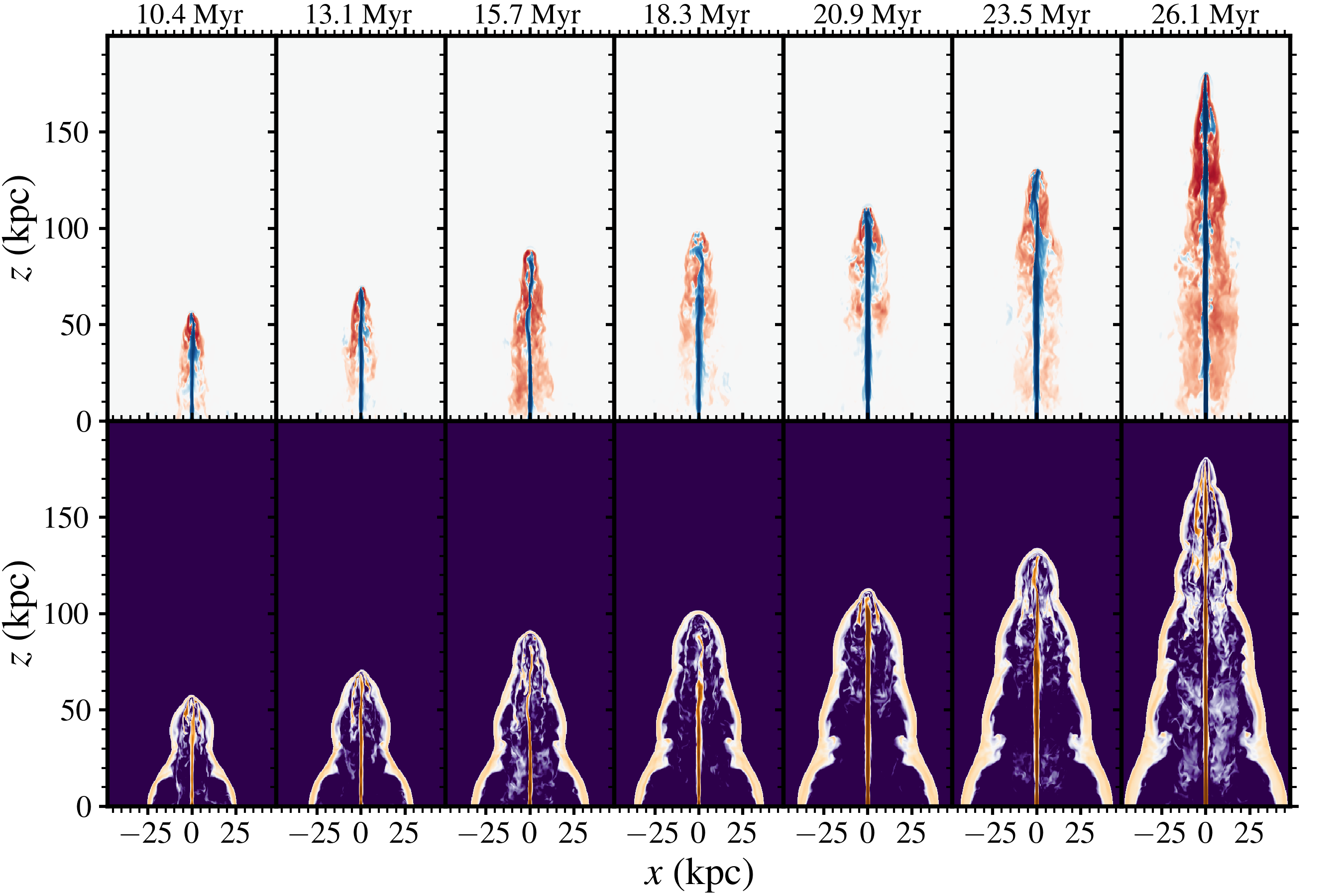}
    \caption{Jet and backflow dynamics in 3D, from a slice through the computational domain at $y=0$. {\em Top:} Vertical velocity component $v_z$. {\em Bottom:} logarithm of the Mach number, $\log_{10}{\cal M}$. Left to right panels show different timestamps at regular intervals. In most cases, fast, supersonic backflow is present and a supersonic jet remains collimated until the edge of the cocoon. However, in some cases, such as in the middle panel, the jet has insufficient ram pressure or thrust and so a `disconnection event' occurs (see text for further discussion). Animated versions of the top (fig12\_vz.mp4) and bottom (fig12\_mach.mp4) panels of this figure are given in the supplementary material. 
    }
    \label{fig:backflow_3d}
\end{figure*}

The jet variability also affects the internal structure of the jet and in particular the location and time evolution of internal and reconfinement shocks within the beam. To illustrate this, we show a `space-time' plot of the logarithmic pressure in the central jet region in Fig.~\ref{fig:spacetime}, where thin vertical profiles are plotted over time at $0.065$~Myr resolution. The plots allow the visualisation of the location of the reconfinement shocks within the jet and also their time evolution, while the envelope of the high pressure region shows the location of the bow shock. The reconfinement shocks can be picked out in the image through stripes in the pressure, which are quasi-periodic along the jet length. However, the characteristic wavelength of these structures changes as the jet varies and the reconfinement shocks typically move upwards as the jet power increases. The appearance of a bright termination shock correlates with periods of fast advance and high power, as expected. The jet variability also leads to propagating internal shocks within the jet beam; a prominent example starts at $7$~Myr and forms a diagonal structure in the space-time plot. Some of the results are similar to those reported by \cite{gomez_hydrodynamical_1997}, who found that velocity perturbations in jet beams could create knot-like structures of synchrotron emission moving at a superluminal apparent velocity. The implications of these internal shocks, which are interesting as sites of particle acceleration, are discussed further in section~\ref{sec:discuss_particle}. 

\section{Observable Properties}
\label{sec:synchro}
\label{sec:obs}
We do not include a detailed model for the synchrotron electrons or magnetic fields in our simulations, but we can nevertheless make an estimate of the synchrotron emission produced by our simulated jets using a pseudo-emissivity. Following a standard approach in the literature, we start by assuming that the energy density of the nonthermal electrons is $U_e$ and the energy density of the magnetic field is related to this by a partioning factor $\eta_b$ such that $U_B=\eta_b U_e$. In this case, for a power-law distribution of nonthermal electrons, it can be shown \cite[e.g.][]{longair_high_1994} that the synchrotron emissivity $j_\nu$ obeys $j_\nu \propto P^{(q+5)/4}$, where $q$ is the electron spectral index. We assume that minimal particle acceleration takes place outside the jet and cocoon material, so multiply the emissivity by the jet tracer $C_{\rm j}$, giving a pseudo-emissivity 
\begin{equation}
    j_{\rm pseudo} = C_{\rm j} P^{(q+5)/4}.
\end{equation}
This quantity must be converted from simulation units to physical units using an emissivity unit $j_0$ that is given by
\begin{equation}
    j_0 = A(q, \gamma_{e,{\rm min}}, \gamma_{e,{\rm max}}) \,\nu^{\frac{q-1}{2}}\,\eta_b^{\frac{q+1}{4}}
    (1+\eta_b)^{-\frac{q+5}{4}}
    \,P_0^{\frac{q+5}{4}} 
\end{equation}
where $A(q, \gamma_{e,{\rm min}}, \gamma_{e,{\rm max}})$ is a function of $q$ and the minimum and maximum Lorentz factors of the electrons, $\gamma_{e,{\rm min}}$ and $\gamma_{e,{\rm max}}$. The full equation in units of specific luminosity per unit solid angle, equal to  $j_0 {\cal L}_0^3 /(4\pi)$, is given by \cite{hardcastle_numerical_2013}. We follow \cite{hardcastle_numerical_2013} in taking $q=2.2$, slightly steeper than the canonical shock acceleration value of $q=2$; this choice is  appropriate for particle acceleration at relativistic shocks \citep{achterberg_particle_2001,kirk_shock_2001}, and also reproduces a fairly typical synchrotron spectral index of $\alpha = (q-1)/2 = 0.6$. We do not include a non-radiating pressure component in our simulations, which is akin to assuming that nonthermal electron acceleration is very efficient and that there is not a dominant hadronic component to the pressure. 

We compute the emissivity and monochromatic luminosity at $144$~{\rm MHz}, for comparison with results from the {\sl Low-Frequency Array} (LOFAR) and in particular the LOFAR Two-metre Sky Survey \citep[LoTSS;][]{shimwell_lofar_2017}, which has provided an unprecedented census of radio emission from AGN jets at these frequencies \citep[e.g.][]{hardcastle_radio-loud_2019}. In addition, our approach, which neglects the impact of synchrotron cooling, is more likely to be a good approximation at these low radio frequencies, because they are probing slower cooling electron populations than, say, $1.4$ or $5$~GHz. 

\subsection{Luminosity-size evolution}

\begin{figure*}
    % \centering
    \includegraphics[width=1.0\columnwidth]{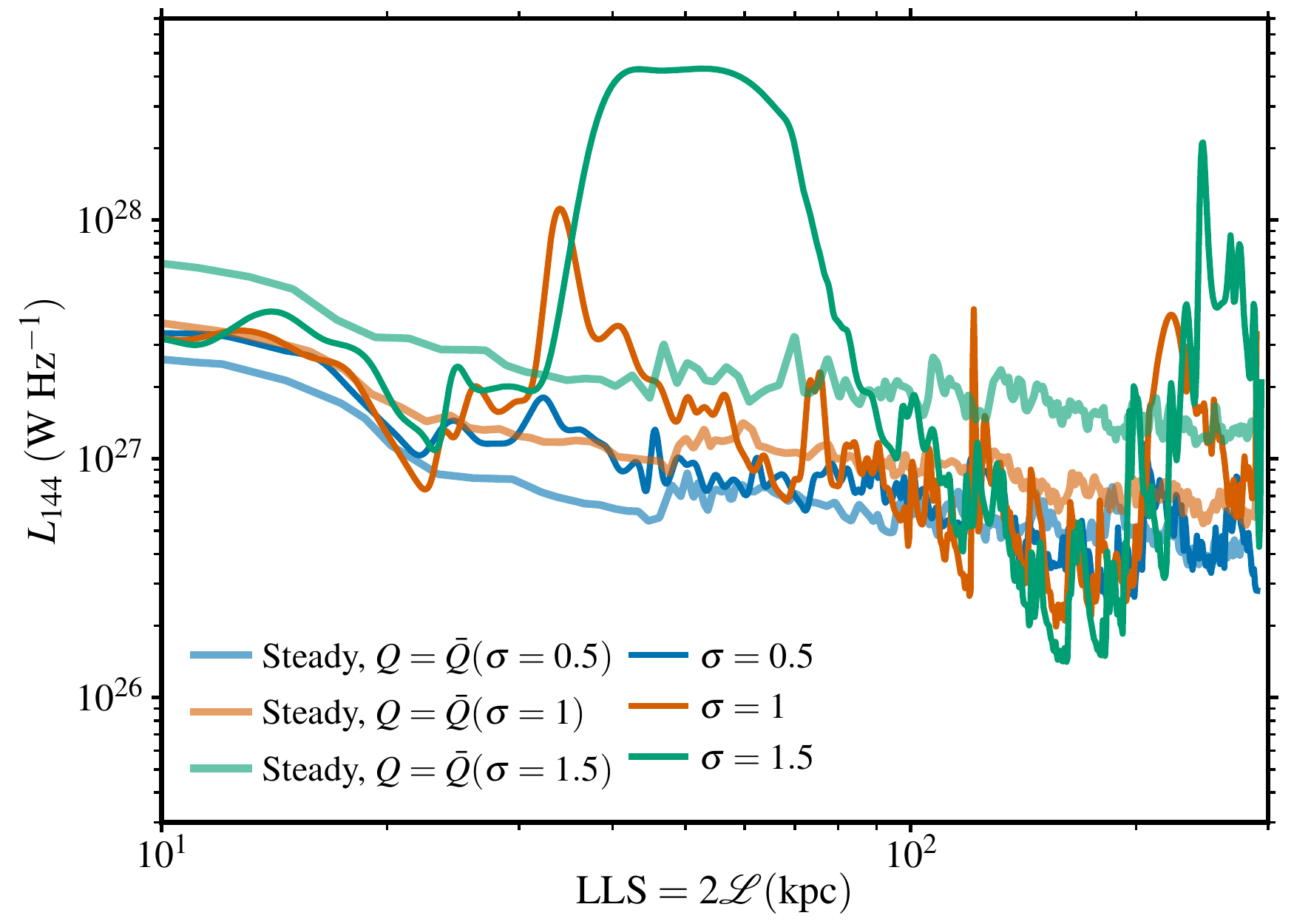}
     \includegraphics[width=1.0\columnwidth]{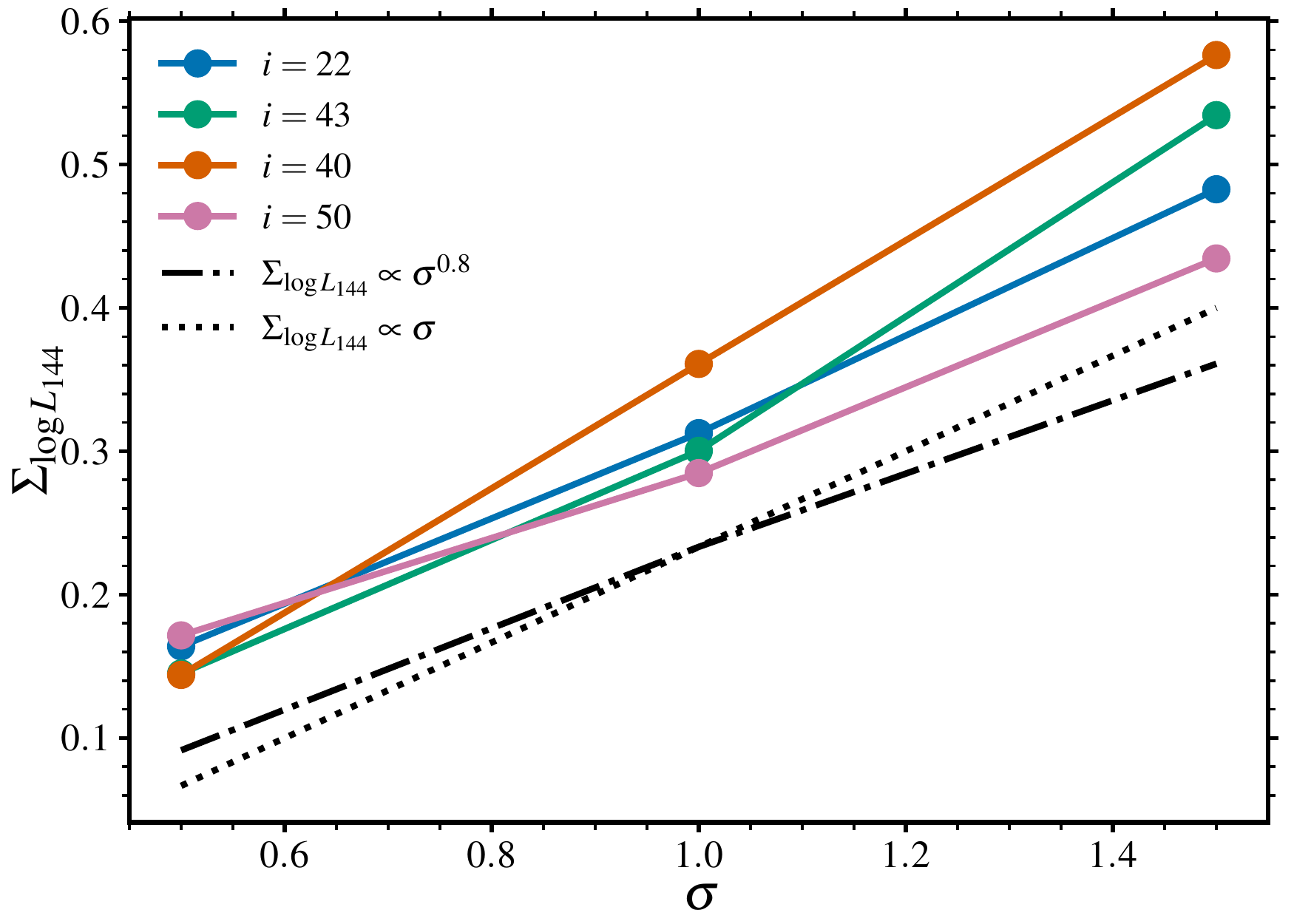}
    \caption{{\sl Left:} Luminosity-size (`$P$-$D$') diagram showing the impact of jet variability. The monochromatic radio luminosity at 144\,MHz is plotted against largest linear size in kpc. Each colour corresponds to a different value of $\protect\sigma$, 
    and each variable jet is compared to a steady jet with the same mean jet power. 
    {\sl Right:} Fractional standard deviation of $\protect\log L_{144}$, 
$\protect\Sigma_{\protect\log L_{144}}$ as a function of $\protect\sigma$, designed to influence the typical `scatter' in radio luminosity for a given level of jet variability. 
For comparison, we show lines of proportionality of $\protect\Sigma_{\protect\log L_{144}} \protect\propto \protect\sigma$ and $\protect\Sigma_{\protect\log L_{144}} \protect\propto \protect\sigma^{0.8}$. 
The dependence is approximately linear and only a modest 
0.2~dex scatter is produced for low $\protect\sigma$ runs. The colours of the lines for the different seeds match those in Figs~ \ref{fig:power_series} and \ref{fig:4panel_sigma1.5}.}
    \label{fig:lum_size}
\end{figure*}

The left-hand panel of Fig.~\ref{fig:lum_size} shows the luminosity-size evolution for three different values of $\sigma$ and a single seed, as compared to steady jets with the same mean jet power ($Q=\overline{Q}(\sigma)$). Unsurprisingly, the jet power variability introduces significant variation in the predicted radio luminosity. To quantify this variation, we can compare the luminosity of the variable jet, $L_{144,{\rm var}}(\sigma,t)$, to the luminosity of the steady jet with the same mean jet power, $L_{144,{\rm steady}}(\sigma,t)$. We then compute the fractional standard deviation in logarithmic space, which we define as
\begin{equation}
\Sigma_{\log L_{144}} (\sigma) = \rm {SD} \left[ \log_{10} \left( \frac{L_{144,{\rm var}}(\sigma,t)} {L_{144,{\rm steady}}(\sigma,t)} \right) \right]
\end{equation}
where ${\rm SD}$ is the standard deviation (calculated across time bins). $\Sigma_{\log L_{144}}$ can be though of as the characteristic scatter in logarithmic luminosity introduced by jet variability with an amplitude $\sigma$. We show $\Sigma_{\log L_{144}}$ as a function of $\sigma$ for each RNG seed in the right-hand panel of Fig.~\ref{fig:lum_size}. We find an approximately linear relation between the two (trend-lines of $\Sigma_{\log L_{144}} \propto \sigma$ and $\Sigma_{\log L_{144}} \propto \sigma^{0.8}$ are shown for comparison). Although the the scatter in radio luminosity compared to the steady jet case can be dramatic, for for $\sigma=0.5$ it is relatively small, around 0.2 dex. Similarly, there is quite a close correspondence between the luminosity-size tracks from a steady jet and the $\sigma=0.5$ simulation. We discuss these results in the context of specific models of AGN fuelling in section~\ref{sec:discuss_fuel}. 

\subsection{Jet power-luminosity relation}
\label{sec:jet_power}
The power of AGN jets is an important quantity for understanding their role in AGN feedback \citep[e.g.][]{dubois_jet-regulated_2010,antognini_lifetime_2012,mukherjee_relativistic_2016} and pinning down the relationship with the accretion disc and black hole \citep[e.g.][]{sikora_radio_2007,blandford_relativistic_2019,davis_magnetohydrodynamic_2020}. Estimating jet power is challenging \cite[see ][]{hardcastle_radio_2020}. \cite{godfrey_agn_2013} estimate jet power based on inferred physical conditions in radio galaxy hotspots, but it is more common to use the so-called `cavity power', or a closely related quantity \citep{cavagnolo_relationship_2010,birzan_radiative_2008,ineson_representative_2017}. The basic approach here is to estimate the total energy or enthalpy in an X-ray cavity and divide by an appropriate timescale for inflation of these cavities, typically a buoyancy time, or advance time. Plotting this inferred power as a function of radio luminosity then gives the jet power to luminosity ($Q-L$) correlation, which can be used to estimate approximate jet powers in sources without X-ray cavity constraints, and also probes the radiative efficiency of the system. 

Flickering variability has implications for this type of jet power estimate, and the $Q-L$ relation, in two main ways. First, it may lead to timescales being under- or over-estimated depending on the true activity time and the ratio with, say, the current advance speed. Second, instantaneous power variability will create corresponding variation in the luminosity and mean power over time which can introduce additional scatter or bias in the power-luminosity relation.

\begin{figure}
    \centering
    \includegraphics[width=\linewidth]{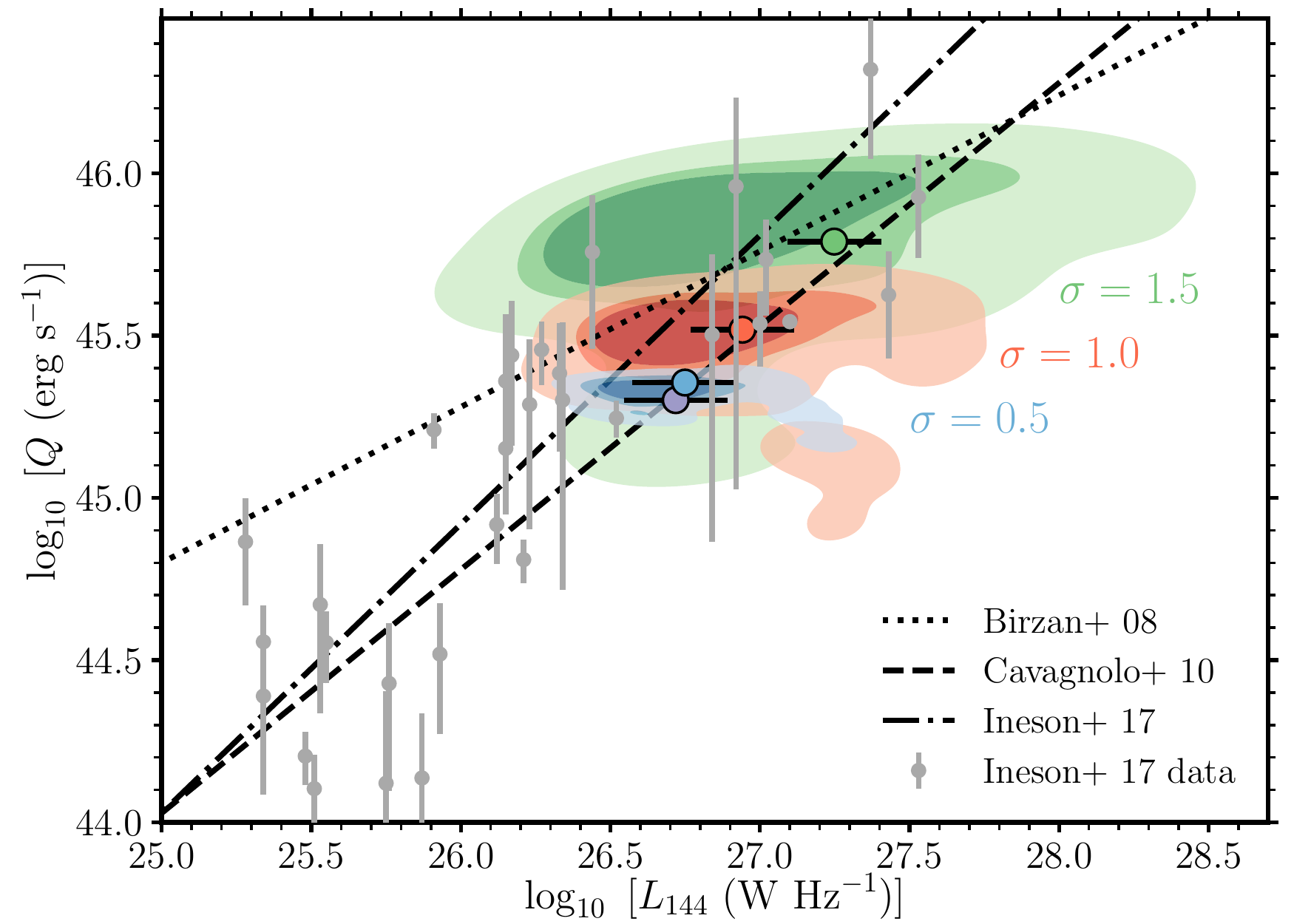}
    \caption{The scatter introduced into the jet power luminosity relation by jet variability. 
    The contours show the kernel density estimate (KDE) containing $50\%$, $68\%$ and $95\%$ of the points, where the KDE is calculated from the luminosity and mean jet power over time throughout a single simulation with $i=22$ and varying $\sigma$. The mean jet power plotted here is not the mean jet power of the entire time series, but is calculated up to time $t$, so is given by $t^{-1}~\int^t_0 Q(t^\prime) dt^\prime$. The different coloured contours therefore illustrate the typical scatter introduced into a jet power-luminosity relation by variability of a given $\sigma$. The coloured circles and errorbars show the mean and standard deviation of the luminosity for steady jets with constant jet powers with the same mean powers as each of the variable simulations, while the various lines show the correlations reported by \protect\cite{birzan_radiative_2008}, \protect\cite{cavagnolo_relationship_2010} and \protect\cite{ineson_representative_2017}.  
    For $\sigma=0.5$, the variability has a limited effect with comparable scatter to the steady jet case, whereas for $\sigma=1.5$ the scatter is dramatic; in both cases the results echo those of Fig.~\ref{fig:lum_size}. 
    }
    \label{fig:power-lum}
\end{figure}

To address the second point concretely, we estimate the scatter introduced using predictions from our 2D simulations. We take the $144~{\rm MHz}$ luminosity from our simulations and calculate the mean jet power at each time $t$, defined as 
\begin{equation}
    \tilde{Q}(t) = \frac{1}{t} \int_0^t Q(t^\prime) dt^\prime \, .
\end{equation}
As $t\to \infty$, $\tilde{Q}(t)\to \overline{Q}(\sigma)$ as defined by equation~\ref{eq:mean}, but at earlier times it depends on the particular time series and exactly when the jet has high or low power episodes. The results are plotted in Fig.~\ref{fig:power-lum}, which shows a kernel density estimate of the results from all of the $i=22$ simulations, for each value of $\sigma$. In addition, the results from steady jets with the same (asymptotic) mean jet powers are shown. For $\sigma=0.5$, the scatter in the relation is comparable to that of the corresponding steady jet, but for more variable jets the scatter is significant. It is also interesting that the steady jet points line up so well with, in particular, the trend-line from \cite{cavagnolo_relationship_2010}. Such a result suggests that the luminosities we are predicting are reasonable, but is also consistent with results from \cite{ineson_representative_2017} and \cite{croston_particle_2018} showing that FRII radio galaxies do not need a large non-radiating pressure component. 

One might invert the question here and ask whether the large scatter in $Q-L_{144}$ space for $\sigma=1.5$ is already inconsistent with the observed scatter -- if so, it would imply that the characteristic variability in the population is smaller than in our most variable simulation. In Fig.~\ref{fig:power-lum} we also plot the data points from the representative sample of FRII radio galaxies conducted by \cite{ineson_representative_2017}. The characteristic scatter in the data is comparable to that from the $\sigma=1$ and $\sigma=1.5$ simulations, showing that the data from this sample do not rule out high levels of variability (although that is not to say they have any causal connection, either). We note here that the data are obtained from relatively mature radio galaxies and so the simulation results might actually be expected to span a wider range of $L_{144}$ and $Q$ than the observational sample. 

\subsection{Ray-traced synthetic images}
\label{sec:images}
\begin{figure*}
    \centering
    \includegraphics[width=1.0\linewidth]{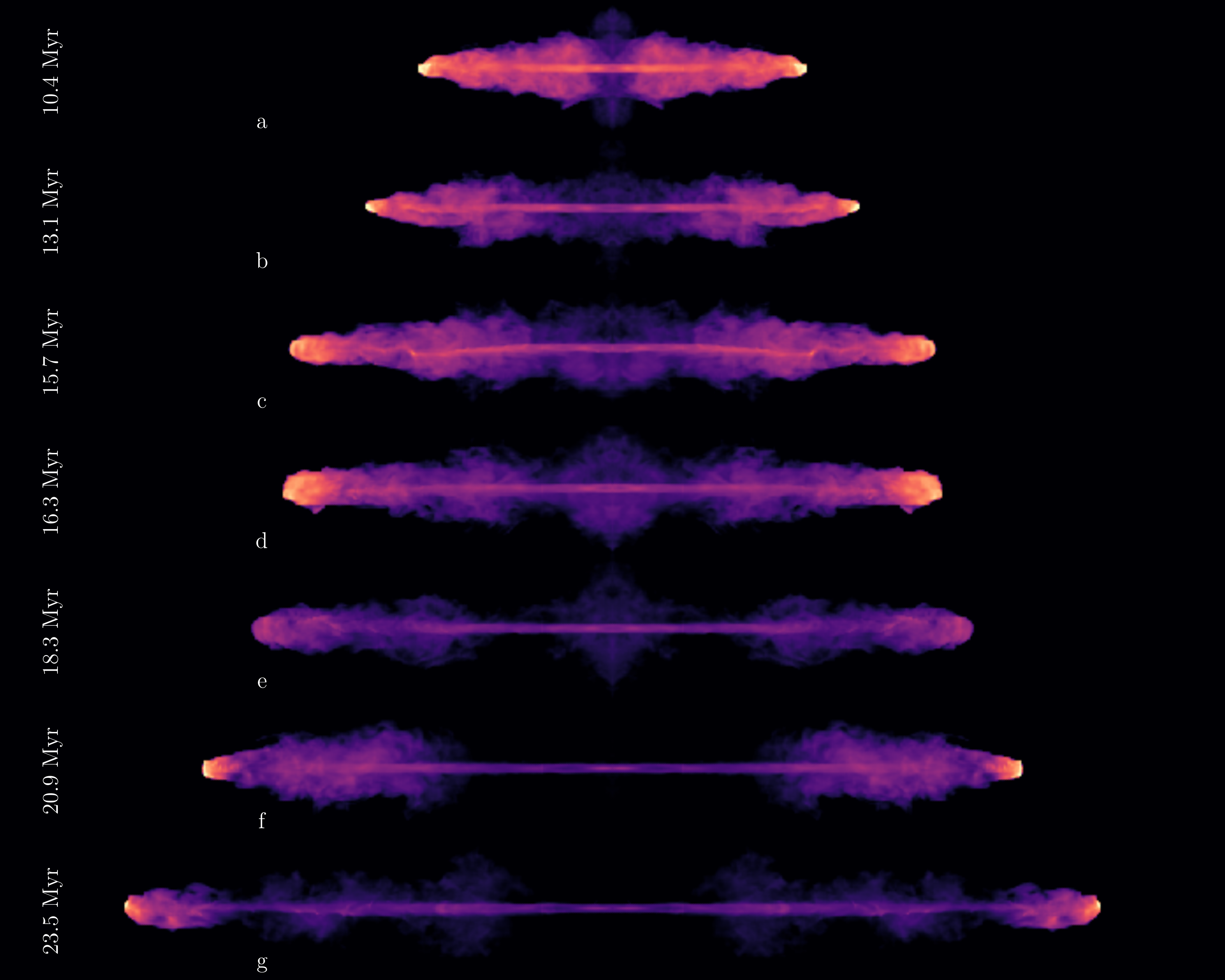}
    \caption{Synthetic radio images calculated from our 3D simulation using the pseudo-emissivity approach defined in section~\ref{sec:synchro}. The ray-traced images are shown for seven different times at regular intervals, and the brightness is plotted on a logarithmic scale with a 3 dex dynamic range. The images are reflected about $z=0$ to mirror the appearance of `classical double' radio galaxies.}
    \label{fig:ray_trace}
\end{figure*}

We close our examination of the observational properties predicted by our simulation by creating synthetic radio images from our fiducial 3D simulation.  To produce synthetic images, we trace rays through the simulation domain assuming optically thin emission for a given observer angle, neglecting any relativistic boosting effects. A selection of images produced using this method is shown in Fig.~\ref{fig:ray_trace}, for seven different time-stamps and fixed polar ($\theta$) and azimuthal ($\varphi$) viewing angles. In each case, the observed image is plotted logarithmically with a dynamic range of three decades and a maximum brightness chosen appropriately for each individual image. 

The first obvious conclusion to be drawn from Fig.~\ref{fig:ray_trace} is that flickering variability can produce changes in observed morphology over the course of a radio galaxy's life. The contrast between the hotspot and the lobes, the length of the backflow/plume region extending backwards from the jet head, and knot-like features along the jet are three examples of properties that evolve. There are a number of specific features in the synthetic images which we highlight:
\begin{itemize}
    \item {\bf Hotspot transience:} Hotspots are not always present in these simulations, driven by the variability of the jet power. High-power states lead to FRII-like morphologies with clear hotspots. In low-power states, not enough power is supplied through the jet, and either the hotspot is dimmed or the jet does not reach the termination region and is instead disrupted (the jet discontinuity effect discussed in section~\ref{sec:3d_dynamics}). 
    \item {\bf `Lobed FR-I' morphology and lobe-hotspot contrast:} The intermittency of hotspots means that the degree of edge-brightening changes - for example, in panel e), the brightness is more uniform across the image compared to panel d), due to a drop in jet power between the two. This change in brightness distribution is interesting for the FRI-FRII dichotomy. At times, the morphology of the system resembles that of a `lobed FR-I' source, in that lobes are present but the clear hotspot typically associated with FRII radio galaxies is not present.
    \item {\bf Jet knots and kinks:} In some cases, knots can be seen along the jet length where internal shocks have formed -- these knots can be both quasi-steady or fast-moving. Knot structures can also be formed close to kinks in the jet, as is particularly apparent in panel c) of Fig~\ref{fig:ray_trace}.
    \item {\bf Illuminated lobe structure:} Structure far from the jet head can be seen in the synthetic images, where the pressure within the lobe is fairly high. These structures are most visible when the dynamic range in brightness is lower, and can correspond to interactions between the cocoon and jet, or other high-pressure regions in the lobe. 
    \item {\bf Multiple hotspots and patchy brightness patterns:} Possible evidence for multiple hotspots or bright regions in the jet head region can be seen in panels c) and d) of Fig.~\ref{fig:ray_trace}. In addition, the lobe structure near the jet head can be rather patchy. 
\end{itemize}
We discuss some of the above points further in sections~\ref{sec:discuss_fuel} and \ref{sec:discuss_obs}. 

\section{Discussion} 
\label{sec:discuss}

% \James{Work in progress.
% HEAVY TAILED distributions of advance speeds? 
% }

\subsection{The link between radio galaxies and AGN fuelling}
\label{sec:discuss_fuel}

\defcitealias{gaspari_raining_2017}{G17}
As described in the introduction, the choice to adopt a flicker noise power spectrum with a log-normal distribution of jet powers here is partly driven by simulations of the fuelling of the central AGN region. 
In particular, the chaotic cold accretion (CCA) model \citep{gaspari2013,gaspari_self-regulated_2016,gaspari_raining_2017} is particularly interesting for our work. \citet[][hereafter \citetalias{gaspari_raining_2017}]{gaspari_raining_2017} study CCA using high-resolution hydrodynamical simulations. \citetalias{gaspari_raining_2017} show that radiative cooling leads to a multiphase structure, causing  what they refer to as a ``top-down condensation cascade'' of $\sim 10^4~{\rm K}$ filamentary structures. The condensing structures eventually reach the central sub-kpc region, and the accretion rate ends up following a log-normal distribution with a flicker noise or $1/f$ power spectrum on $\sim 0.1-10~{\rm Myr}$ timescales. As mentioned in the introduction, this behaviour is indicative of a multiplicative physical process, similar to that observed on shorter timescales in AGN and other accreting systems.

CCA is just one model for how AGN fuelling might proceed, but it is nevertheless instructive to consider how it might imprint itself on AGN jet activity. We do this by directly comparing our considered jet power time series to the accretion rate distribution predicted in the CCA simulations of \citetalias{gaspari_raining_2017}. The distribution of accretion rates reported by \citetalias{gaspari_raining_2017} is slightly skewed, but is well-approximated by a log-normal distribution with $\sigma=1/3$. If we assume that $Q_j \propto \dot{M} c^2$ with constant efficiency -- and ignore the detailed physics of the accretion disc -- this can be compared to the $\sigma$ values we have simulated. Under these crude assumptions, CCA should thus produce variability which is less dramatic than our $\sigma=0.5$ simulation, suggesting luminosity variations compared to steady jets of $\approx 0.2~{\rm dex}$ (Fig.~\ref{fig:lum_size}) and fairly small changes in morphology. Thus, if CCA produces accretion rate modulations of the amplitude suggested by \citetalias{gaspari_raining_2017}, the imprints on the macroscopic radio galaxy properties might be rather inconsequential. In the simulations presented by \cite{yang_how_2016}, the amplitude of variability is fairly similar to that in CCA (we estimate jet power variations at the  $\sigma \approx 0.35$ level from their figure 1), and so similar conclusions apply.  

It is interesting, then, that many radio galaxies do in fact show spectacular evidence of variability in their observed morphologies, as discussed in the introduction. In particular, it seems unlikely that sources like Hercules A, with its distinct ring- or bubble-like structures, can be produced by the relatively gentle flickering induced by chaotic fuelling. Similar principles apply to radio galaxies with clearly separated inner and outer lobe structures, such as Centaurus A, or double-double radio galaxies. It is likely that higher amplitude variability, discrete accretion triggers, or redder noise variability has a role to play in shaping the observed properties of these sources. Indeed, it is possible there is a continuum of noise properties in AGN which depend on the environment, cooling physics, AGN feedback and merger history of the host galaxy. To investigate this, it would be useful to include special relativistic effects in self-consistent fuelling/feedback simulations \citep[e.g.][]{yang_how_2016,beckmann_dense_2019}, and also to predict radio images from these simulations, perhaps in a range of environments. 

\subsection{Other observational implications and tests}
\label{sec:discuss_obs}

In addition to the aforementioned morphological signatures, there are also likely to be more subtle imprints of variability in radio galaxies. Many of the potential spectral signatures, such as spectral hardening beyond a cooling break, or spatially distinct populations of electrons with differing cooling break frequencies or maximum energies, are discussed by \cite{matthews_particle_2021}. In a related study, \cite{maccagni_flickering_2020} find evidence of flickering using spatially resolved synchrotron spectra of radio galaxy Fornax A. As pointed out in section~\ref{sec:images}, variability could also have an impact on multiple hotspot structures, jet knots, and the degree of inhomegenity in brightness, although it is hard to tell if these features are specifically related to variability. With this in mind, we note that \cite{mahatma2023} find patchy brightness and spectral index distributions in regions of the lobes close to the hotspot of 3C 34 and 3C , as well as striking jet knot features in 3C 34 in particular. Multiple hotspots have been observed in a number of radio galaxies, including Cygnus A \citep{williams_multiple_1985,carilli_cygnus_1996,araudo_maximum_2018}, and a recent study by \cite{horton2023} found that jet precession could create a plethora of multiple hotspot phenomena. We have not conducted a full investigation on this topic, but our work suggests that flickering is also important for determining hotspot prevalence and, possibly, multiplicity. 

Jet flickering also has implications for the FR dichotomy. Although there is not a single jet power that sets the divide between FRI and FRII sources given the large overlap in radio luminosity \citep{mingo_revisiting_2019}, one might imagine that -- for a given environment or cluster richness -- there is a critical jet power that basically determines the FR class. In such a scenario, a variable jet power might cause the jet to cross this critical power threshold and transition between morphological classes; in high power states, the jet would remain well-collimated, then collapse and disrupt if the power drops. We have observed behaviour along these lines through the jet discontinuity effect and changes in brightness distribution we discussed in sections~\ref{sec:3d_dynamics} and \ref{sec:images}, respectively. In detail, disruption of the jet depends on the relative growth rates of instabilities in and aroud the jet beam, such as Kelvin-Helmholtz, centrifugal and current-driven kink instabilities \citep[see][for discussions]{appl_current-driven_2000,gourgouliatos_reconfinement_2018,wang2023}, as well as the entrainment of material from the jet's surroundings.  This topic is challenging to investigate given the complexity of the entrainment physics and the difficulties in simulating FRI sources realistically, but it is an interesting avenue for the future. 

There are a few further ways in which evidence of flickering variability might be searched for observationally, beyond the morphological and spectral signatures already discussed here and by \cite{matthews_particle_2021}. A general feature of a flickering jet power is that the instantaneous jet power is different to the time-averaged one, meaning that a comparison of different observational measures of energetics and timescales may prove fruitful. In particular, we can think of a couple of possible experiments. First, one could compare jet powers estimated from the hotspot parameters \citep{godfrey_agn_2013} to those from the lobe energetics. The former will probe the instantaneous jet power, and so the scatter or systematic bias in these estimates can encode variability at the population level. Second, a comparison of the dynamical timescale inferred from current hotspot advance speeds with that inferred from spectral ages or maximum source lifetimes. Such an exercise has been carried out by \cite{kappes_lofar_2019} with LOFAR observations of S5 0836+710, showing that the advance speed must have been significantly higher in the past. While this behaviour could be caused by a steady decrease in jet power or the ambient density profile, the study shows the potential for studying variability through comparison of the important physical timescales or speeds in radio galaxies. 

\subsection{Particle acceleration and ultrahigh energy cosmic rays}
\label{sec:discuss_particle}
Particle acceleration in AGN jets is thought to take place in sites of energy dissipation, specifically shocks, reconnection sites and MHD turbulence (see \citealt{matthews_particle_2020} for a review). compared to a steady jet, variability clearly changes the behaviour of internal and recollimation shocks as well causing the termination shock to be intermittent and the backflow and turbulence to change in character in tandem with the jet power. These changes have a knock on effect on particle acceleration. 

In a recent study, \cite{matthews_particle_2021} studied particle acceleration in variable jets using a simplified (non-hydrodynamic) model for the evolution of the jet-lobe system. UHECRs of a given energy $E$ and charge $Z$ can only be accelerated by astrophysical sources with a kinetic power $Q_k \gtrsim Q_{k,{\rm crit}} = 10^{44}~{\rm erg~s}^{-1}\epsilon_b \eta^2E_{10}^2 Z^{-2}$, where $E_{10} \equiv E/(10~{\rm EeV})$, $\epsilon_b$ is the fraction of energy (e.g. at the shock) contained in the magnetic field, and $\eta$ is a parameter describing how close the particle diffusion is to the optimal (Bohm) regime. \cite{matthews_particle_2021} showed that, in a flickering jet, UHECRs are only accelerated during high-power episodes and that the influence of the UHECR escape time means that the UHECR luminosity over time behaves as a smoothed version of the input luminosity, with the UHECR luminosity only responding to peaks that exceed $Q_{k,{\rm crit}}$. If UHECRs are accelerated in backflow shocks, as suggested by \cite{matthews_ultrahigh_2019}, then this would further favour the production of UHECRs during high-power episodes, because we have shown that strong backflow is more prevalent in these periods. 

It is not just the termination shock and backflow that respond to variability. The lobe conditions are generally more turbulent during high states, so any particle acceleration due to second-order Fermi processes in the lobes \citep[e.g.][]{osullivan_stochastic_2009,hardcastle_high-energy_2009,wykes_mass_2013} would also be expected to be dominated by these episodes. Additionally, as discussed in section~\ref{sec:3d_dynamics}, jet variability creates the possibility of colliding internal shocks. Internal shocks are thought to be efficient particle acceleration sites in gamma-ray bursts \citep{piran_physics_2004} and may explain propagating knot-like features in, e.g., the M87 jet \citep[e.g.][]{bicknell_understanding_1996,spada_internal_2001,bai_radio/x-ray_2003}. Our work suggests internal shocks caused by variability are also important on larger scales in radio galaxies, although distinguishing them observationally from quasi-periodic (in space) and quasi-steady reconfinement shocks may be challenging. 

Finally, we note that we have not included a sub-grid model of particle acceleration in our simulations. There are a number of studies, particularly in recent years, which evolve a population of non-thermal electrons in tandem with MHD simulations \citep[e.g.][]{tregillis2001,tregillis2004,vaidya_particle_2018,mukherjee_new_2020,walg_relativistic_2020,yates-jones_praise_2022,kundu2022,seo_simulation_2023}. While the technique for injecting and evolving the electron populations varies somewhat in these studies, all of them typically involve modelling the particle acceleration at shocks and by turbulence in the jet-cocoon system. In future, such techniques could be readily applied to simulations of flickering jets similar to ours, which would allow a more detailed investigation of how particle acceleration proceeds during different power episodes. 

\subsection{Limitations and missing physics}
\label{sec:limitations}

We have made a number of assumptions in this work which, to some extent, limit the applicability and interpretation of the results. Some of those assumptions are discussed above in the relevant subsections, so here we comment briefly only on the remaining important limitations of our work. 

We have not solved the induction equation and have neglected the influence of magnetic fields. Magnetic fields could modify the morphology of the jet in particular near the jet head where, for example, they can lead to the formation of `nose cones` \citep[e.g.][]{clarke_numerical_1986,komissarov_numerical_1999} and alter the dynamics of the jet and the character of cocoon turbulence \citep[e.g.][]{keppens_extragalactic_2008,gaibler_very_2009}. In addition, the variable jet power might have interesting effects on the stability of the jet, perhaps changing the behaviour of the magnetic kink instability. Such an investigation is beyond the scope of this work but certainly merits further thought. 

In this work, we chose to chiefly focus on the parameters governing the variability of the jet ($\sigma$ and RNG seed), and kept the jet width, $r_{\rm j}$, the median jet power, $Q_0$, and the jet to ambient medium density contrast, $\eta$, fixed. Although the values we adopted were reasonable, all of these choices will affect the results obtained. In particular, adopting a larger $r_{\rm j}$ or $\eta$ would lead to a lower $\gammajet$ for a given $Q_{\rm j}$, pushing the jet into a less relativistic regime and ensuring that relativity-specific results -- such as heavily skewed advance speed distributions -- would be less pronounced. In addition, we chose to only vary the Lorentz factor of the jet to achieve the desired variability in jet power. Our main conclusions are quite general, so are unlikely to be dramatically affected by the decisions we made here, but, like any numerical study, any observational comparisons should be interpreted with these subjective choices in mind. 

We either considered a smoothly varying ambient medium or one with small density perturbations to break symmetry in 3D. In reality, the cluster or group environment the jet propagates into is likely to be turbulent and there may be dense clumps or inhomogeneities. Additionally, asymmetries or more complex density and pressure gradients will alter the propagation and morphology of the jet. Indeed, these factors are likely to be important for explaining some of the more exotic morphologies observed. Density inhomegeneities or complex gradients could in principle mimic some of the effects of variability; for example if the jet propagates more quickly through low density region and slows in high density environments. The jet history is in any case challenging to deduce from single epoch observations, but it is worth noting that the effect of clumps is not degenerate with variability; as the jet propagates through dense regions it should slow and become brighter at fixed power, whereas in our variable jet simulations the jet tends to be brightest during periods of fast advance (when $\gammajet$ is high).  

\section{Conclusions} 
\label{sec:conclusions}

We have conducted relativistic hydrodynamic simulations of AGN jets on kpc-scales, with a jet power that varies according to a flicker noise power spectrum. We generated synthetic jet power time series -- in which the jet bulk Lorentz factor, $\gammajet$, was the time-varying parameter --  with four different random number seeds and three variability parameters, $\sigma$. We ran all these simulations in 2D cylindrical geometry, as well as an additional three simulations of steady jets with jets powers equal to the mean jet powers of the three $\sigma$ values. Additionally, we ran one of the most variable simulations in 3D. The simulations allow us to examine how variable fuelling might affect relativistic jets in a systematic fashion and compare to observations of radio galaxies that show evidence of variable jet activity. Our main findings from this investigations are as follows. 

\begin{enumerate}
    \item We find that the morphology of the bow shock (in particular) and the cocoon are affected by the variability history of the jet, retaining a `memory' of powerful outbursts that lasts late into the jet evolution. We suggest that radio galaxies may be able to act as observational probes of long-term ($\gtrsim {\rm Myr}$ timescale) AGN fuelling. 
    \item We demonstrate the use of {\em sphericity} -- the ratio of the surface area to that of a sphere with the same volume \citep{1935JG.....43..250W} -- as a morphology metric which does not depend on measuring a width at a specific location along the jet. A 2D analogue of this quantity could be also be used for observations.  
    \item The jet advance speed is fast during periods of high activity, but periods of quiescent or low-level activity instead produce passive periods of Sedov-Taylor-like quasi-spherical inflation. The radius evolution can occasionally approach the Sedov-Taylor $t^{2/5}$ scaling, although this is relatively rare in these specific simulations, and likely to be more common if the input noise is redder.
    \item The strongest backflows and most turbulent lobe conditions are found during the periods of highest activity, because this is when the strongest pressure gradient is established between the jet head and lobe. These periods of high activity are therefore likely to dominate the particle acceleration to very-high or ultrahigh energies in these systems. At late times, the backflow is fast and persistent and its character is dictated more by jet-backflow interaction than the macroscopic pressure difference between jet head and cocoon. 
    \item We track the various energy reservoirs over time in our simulation and show that the kinetic energy of the jet is gradually transferred into cocoon internal energy which responds to jet power variation with a slight delay. The energy in the shocked medium responds more slowly still as it is driven by the over-pressured cocoon. We also find that variability can disrupt the close coupling between cocoon and shocked medium energies, the ratio of which is often close to unity in steady jet simulations but is not in this work. 
    \item In 3D, there are a number of interesting morphological features introduced by variability. Variability can change the degree of edge brightening over time and lead to hotspot intermittency. Colliding internal shocks can be produced along the jet, and the variability causes the locations of reconfinement shocks to change over time. We also observe `jet discontinuities', where the jet is no longer able to remain collimated and deliver its power to a hotspot at the end of the jet. We comment on the implications of these discontinuities for the FR dichotomy. 
    \item Using a pseudo-emissivity to estimate radio luminosity over time, we examine the evolution of our simulations in luminosity-size and luminosity-power parameter space. We find that flickering jet variability introduces a fractional standard deviation in radio luminosity ($L_{144}$) which is roughly linear with $\sigma$. For flickering comparable in amplitude to that predicted for the CCA model by \cite{gaspari_raining_2017}, we find only modest scatter $(\lesssim 0.2~{\rm dex})$ in $L_{144}$. Larger variability amplitudes can, unsurprisingly, significantly modify luminosity-size tracks and introduce large scatter in the relation between mean jet power and luminosity.
    \item We produced synthetic radio images from our fiducial 3D simulations, including ray-tracing, for edge-on views. We find that variability can produce changes in observed morphology over the course of a radio galaxy lifetime. Specifically, the flickering variability changes the degree of edge brightening, lobe-hotspot contrast and can create the appearance of lobed FRI-like morphology. In addition, the flickering jet can create propagating internal shock structures, illuminate parts of the lobe far from the jet head, and exacerbate dual hotspot effects and kink structures in the jet-cocoon system. 
    \item We suggest ways to search for evidence of flickering jets. In particular, as well as using morphological and spectral signatures, we propose comparing `instantaneous' estimates of power or advance speed (inferred from the hotspot) to longer-term, calorimetric measures. For example, one could compare jet powers inferred through analysis of the hotspot, and jet powers measured by calculating the energy in the lobes, to gain insights into variability statistics at the population level. 
\end{enumerate}
Overall, our work suggests that flickering, Myr-timescale variabity in AGN jet power is an important factor in dictating the overall morphology and observational appearance of a radio galaxy. Additionally, radio galaxies may be a useful probe of long-term variability in AGN, because they retain a relatively long-term ($\sim 10s$ of Myr) memory of the variability history of the AGN through their dynamics and synchrotron electrons. More generally, our work further demonstrates that variability of jet power introduces rich hydrodynamic and particle acceleration physics which is ripe for future study.  

\section*{Data Availability} 
The data underlying this article are available from the authors on request. The supplementary material is available in a github repository at \url{https://github.com/jhmatthews/flicker-supplement} or in the published journal article.

\section*{Acknowledgements} 
J.H.M acknowledges funding from the Royal Society. H.W.W acknowledges a summer student stipend from the Institute of Astronomy and an STFC studentship. We thank Andrew Taylor, Katie Savard, Debora Sijacki, Vijay Mahatma and Prakriti Pal Choudhury for helpful discussions and suggestions. We would also like to thank Vasily Belokurov and Matt Auger for organising the 2020/2021 part III and summer student programs at the Institute of Astronomy. We gratefully acknowledge the use of the following software packages: \pluto\ \citep{mignone_pluto_2007}, astropy \citep{astropy2013,astropy2018}, matplotlib 2.0.0 \citep{matplotlib}, scipy \citep{2020SciPy-NMeth} and OpenMPI \citep{openmpi}. This work was performed using resources provided by the Cambridge Service for Data Driven Discovery (CSD3) operated by the University of Cambridge Research Computing Service (www.csd3.cam.ac.uk), provided by Dell EMC and Intel using Tier-2 funding from the Engineering and Physical Sciences Research Council (capital grant EP/T022159/1), and DiRAC funding from the Science and Technology Facilities Council (www.dirac.ac.uk).

% Don't change these lines
\bsp	% typesetting comment
\label{lastpage}
\end{document}